\documentclass[twocolumn,eqsecnum,superscriptaddress,floats,prd,nofootinbib]{revtex4}

\usepackage{amssymb,amsmath,amssymb,amsfonts,amsthm,stmaryrd,mathrsfs,physics,bbm}
\usepackage{graphicx}
\usepackage{subfigure}
\usepackage[T1]{fontenc}
\usepackage{enumerate} % advanced enumerate environment
\usepackage{color} % for colored text
\usepackage{graphpap} % for numbered coordinate grid
\usepackage{xcolor}

\usepackage{hyperref} % for HyperTeX cross-referencing

\newtheorem{proposition}{Proposition}

\newcommand{\sut}{{\mathrm{SU(2)}}}

\newcommand{\heff}{{H_{\rm eff}}}

\newcommand{\GZU}{School of Physics, Guizhou University, Guiyang 550025, China}
\newcommand{\BNU}{School of Physics and Astronomy, \mbox{Key Laboratory of Multiscale Spin Physics (Ministry of Education)}, Beijing Normal University, Beijing 100875, China}
\newcommand{\UOE}{Department Physik, Institut f\"ur Quantengravitation, Theoretische Physik III, Friedrich-Alexander-Universit\"at Erlangen-N\"urnberg, Staudtstra{\ss}e 7/B2, 91058 Erlangen, Germany}

\begin{document}

\title{Black holes and covariance in effective quantum gravity: A solution without Cauchy horizons}
\author{Cong Zhang}
\email{cong.zhang@bnu.edu.cn}
\affiliation{\BNU}
\affiliation{\UOE}

\author{Jerzy Lewandowski}
%\email{jerzy.lewandowski@fuw.edu.pl}
\affiliation{Faculty of Physics, University of Warsaw, Pasteura 5, 02-093 Warsaw, Poland}

\author{Yongge  Ma}
\email{mayg@bnu.edu.cn}
\affiliation{\BNU}

\author{Jinsong Yang}
\email{jsyang@gzu.edu.cn}
\affiliation{\GZU}

\begin{abstract}
The issue of general covariance in effective quantum gravity models within the Hamiltonian framework is addressed. The previously proposed equations for the covariance condition in spherically symmetric models are explicitly derived. By solving this equation, a new effective Hamiltonian constraint is obtained, incorporating free functions that can account for quantum gravity effects. The resulting spacetime structure is analyzed by specifying the free functions. Remarkably, in this model, the classical singularity is replaced by a region where the metric asymptotically approaches a Schwarzschild-de Sitter one with negative mass. Thus, this new quantum-corrected black hole model avoids the Cauchy horizons presented typically in previously studied models. The covariant approach is also applicable to matter coupling in the models.
\end{abstract}

\maketitle

%\tableofcontents
\section{Introduction}
It is widely believed that Einstein's general relativity (GR) is not the final theory of spacetime. One significant reason is the inconsistency between the quantum mechanical description of matter and classical spacetime description of gravity \cite{ashtekar2015general}. Another reason is the presence of gravitational singularities, generally appeared in GR \cite{PhysRevLett.14.57}. A significant approach to meet these challenges is search for a consistent theory of quantum gravity (QG)  \cite{Rovelli:1987df,polchinski1998string,Ambjorn:2001cv,Surya:2019ndm}. To explore QG effects, one approach is to treat QG semiclassically  as an effective field theory \cite{Donoghue:2012zc}. In this treatment, the effective gravity is still described by a metric tensor, but the equations of motion should be modified to account for quantum effects.

Various theories of canonical QG are based on the Hamiltonian formulation (see, e.g., \cite{Jha:2022svf} for the Hamiltonian formulation of general relativity).  In the Hamiltonian framework of classical GR, the dynamics are governed by the diffeomorphism and Hamiltonian constraints. Consequently, in 
a semiclassical theory emerged from certain canonical QG, QG effects are expected to be manifested through some modifications to those constraints. However, if the effective spacetime description was valid,  the Hamiltonian formulation would require a 3+1 decomposition of the spacetime. Then the issue of general covariance arises naturally: under what conditions does a given $3+1$ model in the Hamiltonian formulation describe a generally covariant spacetime theory. This has been a long-standing issue open to debate, particularly in the context of effective theories derived from symmetry-reduced models of canonical QG \cite{Tibrewala:2013kba,Bojowald:2015zha,Wu:2018mhg,Bojowald:2019dry,Bojowald:2020unm,Han:2022rsx,Gambini:2022dec,Bojowald:2022zog,Ashtekar:2023cod,Giesel:2023hys,Bojowald:2024beb}. 

General covariance, when translated into the Hamiltonian framework, can be understood by interpreting GR as a totally constrained system with first-class constraints.
 A system of constraints is of first-class if the constraints form a closed algebra under the Poisson bracket, and the theory is totally constrained if the Hamiltonian is simply a linear combination of these constraints. According to the Dirac-Bergmann theory of constrained Hamiltonian systems \cite{Date:2010xr}, first-class constraints correspond to the gauge symmetries of the system. More precisely,  the Poisson bracket of any linear combination of the constraints with a function on phase space gives rise to an infinitesimal gauge transformation of that function.
In the Hamiltonian framework of classical GR, the spacetime metric is a tensor-valued function on the phase space. Therefore, general covariance requires that its infinitesimal gauge transformation generated by the constraints matches its Lie derivative on a spacetime manifold. However, a subtlety arises since the lapse function and shift vector, as the components of spacetime metric, are usually treated as Lagrangian multipliers rather than phase space variables. The meaning of the gauge transformations of the spacetime metric when the lapse function and shift vector are involved will be clarified in Sec. \ref{sec:sec2} (see also \cite{Pons:1996av,Bojowald:2018xxu} for previous discussions). Despite this subtlety, the overall concept of general covariance remains unaffected.  Thus, the concept of general covariance relies on two key elements: the metric and the constraints. To maintain general covariance, these two elements have to be carefully aligned.  In this paper, it is assumed that the spacetime metric is modified into an effective one due to the QG effects, while keeping its form fixed. Based on this effective metric,  the corresponding expression of the constraints can be restricted to ensure that the effective canonical theory remains covariant. Concretely,  we will keep the diffeomorphism constraint in its classical form while leaving the Hamiltonian constraint unknown to incorporate QG effects. This guarantees that the diffeomorphism constraint still generates spatial diffeomorphism transformations. To determine the form of the Hamiltonian constraint, we require the constraint algebra
to remain closed,  mirroring the classical structure with an additional factor accounting for QG effects. The resulting Hamiltonian constraint will be referred to as the effective Hamiltonian constraint. Since QG effects are expected to resolve the classical singularity, the effective Hamiltonian constraint should lead to the formation of regular black holes (BHs) \cite{Lan:2023cvz}. 

In previous works on effective models of canonical QG \cite{Zhang:2021wex, Zhang:2021xoa, Husain:2021ojz, Han:2022rsx, Husain:2022gwp, Giesel:2023hys, Giesel:2023tsj}, one usually introduces
a specific matter field   to partially fix the gauge freedoms of spacetime diffeomorphisms. The Hamiltonian in this preferred gauge is then quantized, yielding an effective Hamiltonian. 
Thus, the covariance issue is somehow avoided by working within a fixed gauge. However, as discussed in \cite{Bojowald:2024beb}, several issues arise within this framework.
First, if one moved away from the preferred gauge, the effective Hamiltonian had to be modified to ensure that the resulting metric remains diffeomorphically equivalent to the one obtained in the fixed gauge. This modified effective Hamiltonian would no longer be directly derived from the same quantization scheme, suggesting that the quantization strategy depends on the choice of gauge. Additionally, the reliance on the matter fields  poses an obstacle to defining a pure-gravity vacuum, as the presence of matter is inherently tied to the gauge-fixing procedure.

In this paper, we will focus on the spherically symmetric vacuum gravity, where, classically, the only nontrivial solutions to Einstein's equations are the Schwarzschild BH and the Kantowski-Sachs universe. By considering the vacuum case, the gauge fixing with matter fields will not be performed, since our aim is to derive effective Hamiltonian constraints that apply uniformly across different gauges. Once the effective Hamiltonian constraints are obtained, matter coupling can be suitably introduced \cite{Zhang:2024khj}.

This paper is organized as follows. In Sec. \ref{sec:sec2}, the notion of general covariance in the spherically symmetric models of classical theory is discussed in details. Based on this notion, the conditions ensuring covariance for the spherically symmetric effective models of QG are proposed in Sec. \ref{sec:covarianceClassical}. In Sec. \ref{sec:covarianceequation}, those conditions are formulated into tractable ``covariance equations'' for the effective Hamiltonian constraint that can describe a covariant model. A new solution to the covariance equations is proposed in Sec. \ref{sec:threesolutions}. In Sec. \ref{sec:structureds3}, the free functions of the solution are fixed, and the causal structure of the resulting spacetime is discussed. The issue of matter coupling in the covariant model is discussed in Sec. \ref{sec:coupledust}. Finally, our results are summarized in Sec. \ref{sec:summary}.  
The codes for some derivations in this paper are available in \cite{numericalResult}.

\section{General covariance in classical theory}\label{sec:sec2}
%In this work, when dealing with indices, we adopt the following conventions: lower latin letters $a,b,\cdots$ run in $\{0,1\}$ and the Greek letters $\rho,\sigma,\cdots$ run in $\{0,1,2,3\}$. 

Let us start with the Hamiltonian formulation of the spherically symmetric GR in connection formalism. The spatial 3-manifold is denoted as $\Sigma=\mathbb X\times\mathbb S^2$ carrying the $\sut$ action, where $\mathbb X$ denotes an 1-D manifold and $\mathbb S^2$ is the 2-sphere.  Let $(x,\theta,\phi)$ denote the coordinates of $\Sigma$ adapted to the $\sut$ action.  The phase space, denoted by $\mathcal P$, comprises of fields $(K_I,E^I)$ $(I=1,2)$ defined on the quotient manifold $\Sigma/\sut\cong \mathbb X$ (see, e.g., \cite{Bojowald:2004af,Bojowald:2005cb,Gambini:2022hxr} for more details of the kinematical structure). The Ashtekar-Barbero variables $(A_a^i,E^b_j)$ (see, e.g., \cite{PhysRevLett.57.2244,BarberoG:1994eia}) read
\begin{equation}\label{eq:symmetricAE}
\begin{aligned}
A_a^i\tau_i\dd x^a&=K_a^i\tau_i\dd x^a+\Gamma_a^i\tau_i\dd x^a,\\
E^a_i\tau^i\partial_a&=E^1\sin\theta\tau_3\partial_x+E^2\sin\theta\tau_1\partial_\theta+E^2\tau_2\partial_\phi,
\end{aligned}
\end{equation}
where $\Gamma_a^i$ is the spin connection compatible with the densitized triad $E^a_i$, and $K_a^i$ is the extrinsic curvature 1-form given by
\begin{equation}\label{eq:symmetricK}
K_a^i\tau_i\dd x^a=K_1\tau_3\dd x+K_2\tau_1\dd\theta+K_2\sin\theta\tau_2\dd\phi.
\end{equation}
Here we adopt the convention $\tau_j=-i\sigma_j/2$ with $\sigma_j$ being the Pauli matrices. 
The non-vanishing Poisson brackets between the phase space variables are
\begin{equation}
\begin{aligned}
\{K_1(x),E^1(y)\}=&2G\delta(x,y),\\
\{K_2(x),E^2(y)\}=&G\delta(x,y),
\end{aligned}
\end{equation}
where $G$ is the gravitational constant.

This system is totally constrained, so that its dynamics is encoded in a set of first-class constraints: the diffeomorphism constraint $H_x$ and the Hamiltonian constraint $H$. They are expressed as
\begin{equation}\label{eq:diffeoCon}
H_x=\frac{1}{2G}\left(-K_1\partial_xE^1+2E^2\partial_xK_2\right),
\end{equation}
and
\begin{equation}\label{eq:clHHHHHH}
\begin{aligned}
H=&-\frac{1}{2G\sqrt{E^1}}\Bigg[E^2+2E^1K_1K_2+E^2(K_2)^2\\
&+\frac{3(\partial_xE^1)^2}{4E^2}-\partial_x\left(\frac{E^1\partial_xE^1}{E^2}\right)\Big].
\end{aligned}
\end{equation}
The Poisson algebra of  the constraints reads
\begin{equation}\label{eq:constraintalgebra}
\begin{aligned}
\{H_x[N^x_1],H_x[N^x_2]\}&=H_x[N^x_1\partial_xN^x_2-N^x_2\partial_xN^x_1],\\
\{H[N],H_x[M^x]\}&=-H[M^x\partial_xN],\\
\{H[N_1],H[N_2]\}&=H_x[S(N_1\partial_xN_2-N_2\partial_xN_1)],
\end{aligned}
\end{equation}
with the structure function $S$ given by 
\begin{equation}
S=\frac{E^1}{(E^2)^{2}},
\end{equation}
where the abbreviation $F[g]=\int_{\Sigma}F(x)g(x)\dd x$ is applied.

For a given vector field $N^x\partial_x$, it is straightforward to check that 
\begin{equation}\label{eq:KEVecCons}
\begin{aligned}
\{K_1\dd x,H_x[N^x]\}&=\mathcal L_{N^x\partial_x}(K_1\dd x),\\
\{E^1,H_x[N^x]\}&=\mathcal L_{N^x\partial_x}E^1,\\
\{K_2,H_x[N^x]\}&=\mathcal L_{N^x\partial_x}K_2,\\
\{E^2\dd x,H_x[N^x]\}&=\mathcal L_{N^x\partial_x}(E^2\dd x),
\end{aligned}
\end{equation}
where $\mathcal L_{N^x\partial_x}$ denotes the Lie derivative with respect to $N^x\partial_x$. Therefore, $H_x[N^x]$ generates the spatial diffeomorphism transformations along $N^x\partial_x$.  In addition, $K_1$ and $E^2$ should be treated as scalar densities with weight $1$ on $\mathbb X$, while $K_2$  and $E^1$ are scalars.

\subsection{Constructing spacetimes from physical states}\label{sec:4Dmetric}
In the phase space $\mathcal P$, the constraint surface $\overline{\mathcal P}$ is defined as the subspace consisting of points of $(K_I,E^I)$ that vanish the constraints $H$ and $H_x$.  For an arbitrary scalar field $\lambda$ and an arbitrary vector field $\lambda^x\partial_x$ on $\mathbb X$, the linear combination $H[\lambda]+H_x[\lambda^x]$, as a function on $\mathcal P$, generates a 1-parameter family of canonical transformations.  For the first-class constraint system, these canonical transformations preserving $\overline{\mathcal P}$ are interpreted as gauge transformations. Applying the gauge transformations associated with all scalar fields $\lambda$ and vector fields $\lambda^x\partial_x$ on a point $(\mathring K_I,\mathring E^I)\in \overline{\mathcal P}$, one gets a set of points lying in $\overline{\mathcal P}$ which is known as the gauge orbit passing through $(\mathring K_I,\mathring E^I)$, 
denoted by $[(\mathring K_I,\mathring E^I)]$ which is a physical state.  

Now let us consider how to construct the spacetime from a physical state $[(\mathring K_I,\mathring E^I)]$. The desired  4-manifold $M$ of spacetime should be equipped with a scalar field $t$ corresponding to the time of certain 3+1 decomposition and a vector field $T^\mu\partial_\mu$ satisfying $T^\sigma\partial_\sigma t=1$, such that the manifold $M$ can be foliated into slices with $t=$constant and each slice is diffeomorphism to $\Sigma$. In the corresponding Hamiltonian formulation, one identifies the slices $t=t_1$ and $t=t_2$ by using the vector flow of $T^\mu\partial_\mu$. Let $\Psi_{t_1,t_2}$ be the identification between the two slices. Then, one can introduce a  family of embeddings $\varphi_{t_o}:\Sigma\to M$ for $t_o\in\mathbb R$ to map $\Sigma$ to the slice $t=t_o$ of $M$ such that 
$$\varphi_{t_1}=\Psi_{t_1,t_2}\circ \varphi_{t_2},\quad \forall\  t_1,t_2. $$
For a given lapse function $N$ and a given shift vector $N^x\partial_x$, which are allowed to be phase space dependent, one may solve the Hamilton's equations 
\begin{equation}\label{eq:hamiltonEqs}
\begin{aligned}
\dot K_I=&\{K_I,H[N]+H_x[N^x]\},\\
\dot E^I=&\{E^I,H[N]+H_x[N^x]\},
\end{aligned}
\end{equation}
and obtain  a curve $t\mapsto (K_I(t),E^I(t))$ lying in the gauge orbit in $\overline{\mathcal P}$ corresponding to a physical state $[(\mathring K_I,\mathring E^I)]$.  Since $(K_I(t_o),E^I(t_o))$  for each moment $t_o$ are actually fields on $\Sigma$, they are pushed forward to the slice $t=t_o$ of $M$ by the mapping $\varphi_{t_o}$, so do the $N(t_o)$ and $N^x(t_o)\partial_x$. Thus, all these fields have become 4-D objects on $M$. Then we can define the metric $g_{\rho\sigma}$ on $M$ as
\begin{equation}\label{eq:metric}
\begin{aligned}
g_{\rho\sigma}\dd x^\rho\dd x^\sigma=&-N^2\dd t^2+\frac{(E^2)^2}{E^1}(\dd x+N^x\dd t)^2+E^1\dd\Omega^2,
\end{aligned}
\end{equation}  
where $\dd\Omega^2=\dd\theta^2+\sin^2\theta\dd\phi^2$ is the standard line element on $\mathbb S^2$.

It is obvious that the metric \eqref{eq:metric} depends on the choice of $N$ and $N^x\partial_x$. However, as known from classical GR, different metrics constructed from different $N$ and $N^x\partial_x$ are the same up to  4-D diffeomorphism transformations on $M$, implying the covariance of the theory with respect to the metric \eqref{eq:metric}. 
We are going to explore the precise meaning of the covariance in the context of the Hamiltonian formulation.

\subsection{Covariance in Hamiltonian formulation}\label{sec:covariancecl}
In the procedure described in  Sec. \ref{sec:4Dmetric}, the solution of $K_I(t)$ and $E^I(t)$ plays two roles. First, it gives a curve $t\mapsto (K_I(t),E^I(t))$ in the phase space $\mathcal P$. Second, it is mapped onto the 4-D manifold $M$ as 4-D fields. Treating them as 4-D fields on $M$, the Hamilton's equation \eqref{eq:hamiltonEqs} can be written in terms of Lie derivative as  
\begin{equation}\label{eq:effHamton2cl}
\begin{aligned}
\mathcal L_{\mathfrak N}E^1&=\{E^1,H[N]\},\\
\mathcal L_{\mathfrak N}(E^2\dd x\wedge\dd t)&=\{E^2,H[N]\}\dd x\wedge \dd t,\\
\mathcal L_{\mathfrak N}(K_1\dd x\wedge \dd t)&=\{K_1,H[N]\}\dd x\wedge \dd t,\\
\mathcal L_{\mathfrak N}K_2&=\{K_2,H[N]\},
\end{aligned}
\end{equation} 
with $\mathfrak N^\rho\partial_\rho\equiv T^\rho\partial_\rho-N^x\partial_x$. Note that in Eq.~\eqref{eq:effHamton2cl} $E^2$ and $K_1$ are lifted as 2-forms  $E^2  \dd x \wedge \dd t$ and $K_1 \dd x \wedge \dd t$ on $M$, respectively. Equivalently, they can be interpreted as scalar densities of weight 1 on the 2-D manifold $M/\sut\ni(t,x)$.  Shall we treat $E^2$  and  $K_1$ as 1-forms $E^2 \dd x$ and $K_1 \dd x$ on $M$, their Poisson brackets with $H[N]$ would not be the desired Lie derivatives.

Next, we consider an infinitesimal gauge transformation generated by $H[\alpha N]+H_x[\beta^x]$, where $\alpha$ is a scalar field and $\beta^x\partial_x$ is a vector field. It 
will transform the curve $t\mapsto (K_I(t),E^I(t))$ into another curve
\begin{equation}\label{eq:newcurvecl}
t\mapsto (K_I(t)+\epsilon\delta K_I(t),E^I(t)+\epsilon \delta E^I(t)),
\end{equation}
where
\begin{equation}\label{eq:two13}
\begin{aligned}
\delta K_I:=&\{K_I,H[\alpha N]+H_x[\beta^x]\},\\
\delta E^I:=&\{E^I,H[\alpha N]+H_x[\beta^x]\}.
\end{aligned}
\end{equation}
The curve \eqref{eq:newcurvecl} satisfies also the Hamilton's equation \eqref{eq:effHamton2cl} with respect to some new lapse function $N+\epsilon\delta N$ and shift vector $(N^x+\epsilon\delta N^x)\partial_x$, and hence we get
 \begin{equation}\label{eq:eqtocheckEffcl}
\begin{aligned}
&\mathcal L_{\mathfrak N}\delta X+\mathcal L_{\delta \mathfrak N}X\\
=&\{X,H[\delta N]\}+\{\{X,H[N]\},H[\alpha N]+H_x[\beta^x]\},
\end{aligned}
\end{equation}
with the abbreviation $X\equiv K_I,E^I$. By direct calculations, we have
\begin{equation}\label{eq:ldx1cl}
\begin{aligned}
\mathcal L_{\mathfrak N}\delta X=&\{\{X,H[\alpha N]+H_x[\beta^x]\},H[N]\}\\
&+\left\{X,H\big[\mathcal L_{\mathfrak N}(\alpha N)-\{\alpha N,H[N]\}\big]\right\}\\
&+\left\{X,H_x\big[(\mathcal L_{\mathfrak N}\beta )^x-\{\beta^x,H[N]\}\big]\right\}\\
=&-\{\{H[\alpha N]+H_x[\beta^x],H[N]\},X\}\\
&+\{\{X,H[N]\},H[\alpha N]+H_x[\beta^x\}\\
&+\left\{X,H\big[\mathcal L_{\mathfrak N}(\alpha N)-\{\alpha N,H[N]\}\big]\right\}\\
&+\left\{X,H_x\big[(\mathcal L_{\mathfrak N}\beta )^x-\{\beta^x,H[N]\}\big]\right\},
\end{aligned}
\end{equation}
where $\alpha$, $\beta^x$, $N$ and $N^x$ could be phase space dependent. Note that in the second equality of Eq.~\eqref{eq:ldx1cl}, the term $\{\{X,H[\alpha N]+H_x[\beta^x]\},H[N]\}$ gives the Lie derivatives of the phase-space variables in $\{X,H[\alpha N]+H_x[\beta^x]\}$ while the Lie derivatives of the phase-space-independent quantities are included in the other terms. 

By using the constraint algebra \eqref{eq:constraintalgebra}, we get
\begin{equation}\label{eq:twoonesix}
\begin{aligned}
&\{H[\alpha N]+H_x[\beta^x],H[N]\}\\
=&-H_x[ SN^2\partial_x\alpha]+H\big[\{\alpha N,H[N]\}\big]\\
&+H[\beta^x\partial_xN]+H_x\big[\{\beta^x,H[N]\}\big].
\end{aligned}
\end{equation} 
Substituting Eq.~\eqref{eq:twoonesix} into Eq.~\eqref{eq:ldx1cl} results in
\begin{equation}
\begin{aligned}
&\mathcal L_{\mathfrak N}\delta X-\{\{X,H[N]\},H[\alpha N]+H_x[\beta^x\}\\
=&\{H_x[SN^2\partial_x\alpha],X\}-\{H[\beta^x\partial_xN],X\}\\
&+\left\{X,H\big[\mathcal L_{\mathfrak N}(\alpha N)\big]\right\}+\left\{X,H_x\big[(\mathcal L_{\mathfrak N}\beta)^x\big]\right\}.
\end{aligned}
\end{equation}
Thus, Eq.~\eqref{eq:eqtocheckEffcl} can be simplified as
\begin{equation}\label{eq:eqtocheckEff2cl}
\begin{aligned}
&\mathcal L_{\delta \mathfrak N}X+\{H_x[ SN^2\partial_x\alpha ],X\}-\{H[\beta^x\partial_xN],X\}\\
&+\left\{X,H\big[\mathcal L_{\mathfrak N}(\alpha N)\big]\right\}+\left\{X,H_x\big[(\mathcal L_{\mathfrak N}\beta)^x\big]\right\}\\
=&\{X,H[\delta N]\}.
\end{aligned}
\end{equation}
The definition $\mathfrak N^\sigma\partial_\sigma=T^\sigma\partial_\sigma -N^x\partial_x$ implies $\delta \mathfrak N^\sigma\partial_\sigma t=0$. Hence we obtain
\begin{equation}\label{eq:two19}
\mathcal L_{\delta \mathfrak N}X=\{X,H_x[\delta\mathfrak N^x]\}-H_x\big[\{X,\delta \mathfrak N^x\}\big].
\end{equation} 
Plugging Eq.~\eqref{eq:two19} into Eq.~\eqref{eq:eqtocheckEff2cl}, we have
\begin{equation}\label{eq:eqtocheckEff4cl}
\begin{aligned}
&\left\{X,H_x\Big[\delta\mathfrak N^x-E^1(E^2)^{-2}N^2\partial_x\alpha+(\mathcal L_{\mathfrak N}\beta)^x\Big]\right\}\\
=&\left\{X,H\Big[\delta N-\beta^x\partial_xN-\mathcal L_{\mathfrak N}(\alpha N)\Big]\right\}+H_x\big[\{X,\delta \mathfrak N^x\}\big].
\end{aligned}
\end{equation}
Thus implies 
\begin{equation}\label{eq:dNNcl}
\begin{aligned}
\delta N^x\approx &-N^2 S\partial_x\alpha-(\mathcal L_{\beta}\mathfrak N)^x,\\
\delta N\approx &\mathcal L_{\alpha \mathfrak N+\beta  }N+N\mathfrak N^\rho\partial_\rho\alpha,
\end{aligned}
\end{equation}
where the equality $\delta N^x=-\delta \mathfrak N^x$ is used, and the convention $A\approx B$ indicates that $A$ is equal to $B$ on the constraint surface $\overline{\cal P}$.

Since the Hamiltonian constraint $H$ in Eq.~\eqref{eq:clHHHHHH} is independent of the derivatives of $K_I$,  it is easy to obtain
\begin{equation}\label{eq:dE2cl}
\{E^I(x),H[\alpha N]\}\approx \alpha(x)\{E^I(x),H[N]\}.
\end{equation}
Plugging Eq.~\eqref{eq:dE2cl} into Eq.~\eqref{eq:two13}, we get
\begin{equation}\label{eq:dE2pcl}
\begin{aligned}
\delta E^1\approx &\mathcal L_{\alpha \mathfrak N+\beta}E^1,\\
\delta E^2\approx & \mathcal L_{\alpha \mathfrak N+\beta }E^2-E^2\mathfrak N^\mu\partial_\mu \alpha,
\end{aligned}
\end{equation}
where $E^1$ is a scalar while $E^2$ is a scalar density of weight $1$.

By Eqs.~\eqref{eq:dNNcl} and \eqref{eq:dE2pcl}, the infinitesimal gauge transformation for $g_{\rho\sigma}$ generated by $H[\alpha N]+H_x[\beta^x]$ reads
\begin{equation}\label{eq:deltagLcl}
\begin{aligned}
\delta g_{\rho\sigma}\dd x^\rho\dd x^\sigma\approx\mathcal L_{\alpha\mathfrak N+\beta}(g_{\rho\sigma}\dd x^\rho\dd x^\sigma),
\end{aligned}
\end{equation}
which corresponds to its diffeomorphism transformation generated by the vector field $\alpha\mathfrak N+\beta$.  This verifies the covariance of the theory with respect to the metric $g_{\rho\sigma}$.

\section{Conditions of general covariance for effective models}\label{sec:covarianceClassical}

Consider an arbitrary spherically symmetric effective model of some canonical QG theory. We assume that the diffeomorphism constraint keeps the same as the classical expression \eqref{eq:diffeoCon}, such that it still generates spatial diffeomorphism transformations. However, due to QG effects, the effective Hamiltonian constraint  $\heff$ may deviate from the classical one.
In addition, the constraint algebra is assumed to be
\begin{equation}\label{eq:effconstraintalgebra}
\begin{aligned}
\{H_x[N^x_1],H_x[N^x_2]\}&=H_x[N^x_1\partial_xN^x_2-{N^x_2\partial_xN^x_1}],\\
\{\heff[N],H_x[M^x]\}&=-\heff[M^x\partial_xN],\\
\{\heff[N_1],\heff[N_2]\}&=H_x[\mu S(N_1\partial_xN_2-N_2\partial_xN_1)],
\end{aligned}
\end{equation}
where the modification factor $\mu$, as some function of $K_I$ and $E^I$, comes from QG effects. It is then nature to ask the question whether the effective model is still covariant with respect to the metric \eqref{eq:metric}. If it is not, is it possible to find an effective metric with respect to which the effective model is covariant?

Since the effective  model is still a totally constrainted system, we can follow the procedure described in Sec. \ref{sec:4Dmetric} to check its covariance.  As done in Eq.~\eqref{eq:effHamton2cl}, the Hamilton's equation \eqref{eq:hamiltonEqs}, with replacing $H[N]$ by $\heff[N]$, can be written in terms of Lie derivative as  
\begin{equation}\label{eq:effHamton2}
\begin{aligned}
\mathcal L_{\mathfrak N}E^1&=\{E^1,\heff[N]\},\\
\mathcal L_{\mathfrak N}E^2&=\{E^2,\heff[N]\},\\
\mathcal L_{\mathfrak N}K_1&=\{K_1,\heff[N]\},\\
\mathcal L_{\mathfrak N}K_2&=\{K_2,\heff[N]\},
\end{aligned}
\end{equation} 
by the computation  on the 2-D manifold $M/\sut$.

Next, as shown in Sec. \ref{sec:4Dmetric}, the infinitesimal gauge transformation generated by $\heff[\alpha N]+H_x[\beta^x]$ transforms the curve $t\mapsto (K_I(t),E^I(t))$ into another curve
\begin{equation}\label{eq:newcurve}
t\mapsto (K_I(t)+\epsilon\delta K_I(t),E^I(t)+\epsilon \delta E^I(t)),
\end{equation}
with 
\begin{equation}
\begin{aligned}
\delta K_I:=&\{K_I,\heff[\alpha N]+H_x[\beta^x]\},\\
\delta E^I:=&\{E^I,\heff[\alpha N]+H_x[\beta^x]\}.
\end{aligned}
\end{equation}
The curve \eqref{eq:newcurve} satisfies the Hamilton's equation \eqref{eq:effHamton2} with respect to some new lapse function $N+\epsilon\delta N$ and shift vector $(N^x+\epsilon\delta N^x)\partial_x$. Following the derivations for \eqref{eq:dNNcl}, we get
\begin{equation}\label{eq:dNN}
\begin{aligned}
\delta N^x\approx &-N^2\mu S\partial_x\alpha-(\mathcal L_{\beta}\mathfrak N)^x,\\
\delta N\approx &\mathcal L_{\alpha \mathfrak N+\beta  }N+N\mathfrak N^\rho\partial_\rho\alpha,
\end{aligned}
\end{equation}
where, in comparison with Eq.~\eqref{eq:dNNcl}, the factor $\mu$ is involved in the expression of $\delta N^x$.

Since in the effective model $\heff$ could depends on derivatives of $K_1$, we have
\begin{equation}\label{eq:dE2}
\begin{aligned}
&\{E^1(x),\heff[\alpha N]\}\\
=&\alpha(x)\{E^1(x),\heff[N]\}-\Delta_1\\
&+\int \dd y \heff(y)N(y)\{E^1(x),\alpha(y)\},
\end{aligned}
\end{equation}
where
\begin{equation*}
\Delta_1=\sum_{n\geq 1}(-1)^n\sum_{m=1}^n\binom{n}{m}(\partial_x^m\alpha)\frac{\partial^n \left( N(x)\frac{\partial \heff(x)}{\partial(\partial_x^nK_1(x))}\right)}{\partial x^{n-m}}.
\end{equation*}
Equation \eqref{eq:dE2} indicates 
\begin{equation}\label{eq:dE2p}
\delta E^1\approx \mathcal L_{\alpha \mathfrak N+\beta}E^1-\Delta_1.
\end{equation}
Similarly, we have
\begin{equation}\label{eq:dE1}
\begin{aligned}
\delta E^2\approx &\alpha\mathcal L_{\mathfrak N}E^2+\mathcal L_{\beta}E^2-\Delta_2\\
=&\mathcal L_{\alpha \mathfrak N+\beta }E^2-E^2\mathfrak N^\mu\partial_\mu \alpha-\Delta_2,
\end{aligned}
\end{equation}
where
\begin{equation*}
\begin{aligned}
\Delta_2=&\alpha(x)\{E^2(x),\heff[N]\}-\{E^2(x),\heff[\alpha N]\}\\
=&\sum_{n\geq 1}(-1)^n\left[\sum_{m=1}^n\binom{n}{m}(\partial_x^m\alpha)\frac{\partial^n\left( N(x)\frac{\partial \heff(x)}{\partial(\partial_x^nK_2(x))}\right)}{\partial x^{n-m}}\right].
\end{aligned}
\end{equation*}

By Eqs.  \eqref{eq:dNN}, \eqref{eq:dE2p} and \eqref{eq:dE1}, the infinitesimal gauge transformation for $g_{\rho\sigma}$ generated by $\heff[\alpha N]+H_x[\beta^x]$ reads
\begin{equation}\label{eq:deltagL}
\begin{aligned}
&\delta g_{\rho\sigma}\dd x^\rho\dd x^\sigma\\
\approx&\mathcal L_{\alpha\mathfrak N+\beta}(g_{\rho\sigma}\dd x^\rho\dd x^\sigma)\\
&+\left(\frac{\Delta_1}{(E^1)^2}-\frac{2\Delta_2 }{ E^1 E^2}\right)(\dd x+N^x\dd t)^2-\Delta_1\dd\Omega^2\\
&+N^2(1-\mu)(\partial_x\alpha)(2\dd x\dd t+2N^x  (\dd t)^2),
\end{aligned}
\end{equation}
where  the metric $g_{\rho\sigma}$ is given by Eq.~\eqref{eq:metric}. Therefore, in general the theory is no longer covariant with respect to $g_{\rho\sigma}$. 

Comparing Eq.~\eqref{eq:deltagL} with Eq.~\eqref{eq:deltagLcl}, three terms arise to break the desired covariance. While the terms proportional to $\Delta_1$ and $\Delta_2$ could be eliminated by requiring that $\heff$ is independent of the derivatives of $K_I$ as in the classical case, to remove the term proportional to $1-\mu$ is challenging. This motivates us to modify $g_{\rho\sigma}$ into some effective metric $g_{\rho\sigma}^{(\mu)}$, with respect to which the theory is expected to restore its covariance. Following the arguments in \cite{Bojowald:2011aa,Zhang:2024khj}, it is assumed that the effective metric takes the form 
\begin{equation}\label{eq:effg}
\begin{aligned}
g^{(\mu)}_{\rho\sigma}\dd x^\rho\dd x^\sigma=&- N^2\dd t^2+\frac{(E^2)^2}{\mu E^1}(\dd x+N^x\dd t)^2+E^1\dd\Omega^2.
\end{aligned}
\end{equation}
Here, by taking the structure function $\mu S$ in  \eqref{eq:effconstraintalgebra}  as the $(x,x)$-components of the inverse spatial effective metric, the algebra \eqref{eq:effconstraintalgebra}  continues to describe hyper-surface deformations in the same way as in the classical case.
Then, by Eqs.  \eqref{eq:dNN}, \eqref{eq:dE2} and \eqref{eq:dE1}, we get 
\begin{equation}\label{eq:thisresult1p}
\begin{aligned}
&\delta g^{(\mu)}_{\rho\sigma}\dd x^\rho\dd x^\sigma\\
\approx &\mathcal L_{\alpha\mathfrak N+\beta}(g_{\rho\sigma}^{(\mu)}\dd x^\rho\dd x^\sigma)\\
&+\frac{\Delta_1}{\mu (E^1)^2}(\dd x+N^x\dd t)^2-\Delta_1\dd\Omega^2\\
&+\frac{1}{\mu^2 E^1}\left(\mathcal L_{\alpha \mathfrak N+\beta}\mu-\delta\mu-\frac{2\mu\Delta_2}{ E^2}\right)(\dd x+N^x\dd t)^2.
\end{aligned}
\end{equation}
It should be noted that the weak equality $\approx$ in the above equation would be replaced by the strong equality $=$ if $\alpha$, $\beta$, $N$ and $N^x$ were phase space independent. Reversely,  the derivations of Eqs.~\eqref{eq:dNN}, \eqref{eq:dE2p} and \eqref{eq:dE1} indicates that, if one had the identity
\begin{equation}\label{eq:thisresult1}
\begin{aligned}
&\delta g^{(\mu)}_{\rho\sigma}\dd x^\rho\dd x^\sigma\\
=& \mathcal L_{\alpha\mathfrak N+\beta}(g_{\rho\sigma}^{(\mu)}\dd x^\rho\dd x^\sigma)\\
&+\frac{\Delta_1}{\mu (E^1)^2}(\dd x+N^x\dd t)^2-\Delta_1\dd\Omega^2\\
&+\frac{1}{\mu^2 E^1}\left(\mathcal L_{\alpha \mathfrak N+\beta}\mu-\delta\mu-\frac{2\mu\Delta_2}{ E^2}\right)(\dd x+N^x\dd t)^2
\end{aligned}
\end{equation}
for all phase space independent quantities of $\alpha$, $\beta$, $N$ and $N^x$, Eq.~\eqref{eq:thisresult1p} would be satisfied automatically. Therefore, rather than  Eq.~\eqref{eq:thisresult1p}, we only need to consider  Eq.~\eqref{eq:thisresult1} for all phase space independent quantities of $\alpha$, $\beta$, $N$ and $N^x$. Therefore, in order to restore the covariance, i.e., $\delta g^{(\mu)}_{\rho\sigma}\dd x^\rho\dd x^\sigma= \mathcal L_{\alpha\mathfrak N+\beta}(g_{\rho\sigma}^{(\mu)}\dd x^\rho\dd x^\sigma)$, the necessary and sufficient conditions are figured out as follows:
\begin{itemize}
\item[(i)] $\Delta_1=0$ for all phase space independent $\alpha$ and $N$, which is equivalent to require that $\heff$ is independent of the derivatives of $K_1$;
\item[(ii)] the following equation is satisfied for all phase space independent $\alpha$, $\beta$, $N$ and $N^x$:
\begin{equation}\label{eq:dmuLmu}
\mathcal L_{\alpha \mathfrak N+\beta}\mu-\delta\mu-\frac{2\mu\Delta_2}{ E^2}= 0.
\end{equation}
\end{itemize}
In Eq.~\eqref{eq:dmuLmu}, the strong equality $=$ is used instead of the weak equality because the resulting $\mu$ is expected to remain valid for models with matter coupling. More explicitly, if one uses the weak equality $\approx$ here, the expression for $\mu$ could be different for different matter coupling models, because the total constraints for different models with matter coupling are different.  

The Hamilton's equation indicates 
\begin{equation}
\mathcal L_{\mathfrak N}\mu=\{\mu,\heff[N]\}.
\end{equation}
Hence, taking account of the fact that $\mu$ is a spacetime scalar, one has
\begin{equation}
\mathcal L_{\alpha\mathfrak N+\beta}\mu=\alpha \{\mu,\heff[N]\}+\{\mu,H_x[\beta^x]\}.
\end{equation}
Since $\delta \mu$ is defined by
\begin{equation}
\delta\mu=\{\mu,\heff[\alpha N]+H_x[\beta^x]\},
\end{equation}
Eq.~\eqref{eq:dmuLmu} can be simplified as
\begin{equation}\label{eq:conditoiniio1}
\alpha\left\{\frac{\mu}{(E^2)^2},\heff[N]\right\}=\left\{\frac{\mu}{(E^2)^2},\heff[\alpha N]\right\}.
\end{equation}
Since $\Delta_1=0$, Eq.~\eqref{eq:conditoiniio1} is equivalent to 
\begin{equation}\label{eq:conditoiniio}
\alpha\left\{\mu S,\heff[N]\right\}=\left\{\mu S,\heff[\alpha N]\right\},
\end{equation}
where $\mu S$ is the structure function appearing in \eqref{eq:effconstraintalgebra}. 
We now reach the following proposition \footnote{While finalizing this paper, we became aware that Eq.~\eqref{eq:mucond} was independently derived in Ref. \cite{Bojowald:2023xat}, though with an on-shell condition that we deliberately omit in Eq.~\eqref{eq:mucond}. It should be noted that the on-shell condition in the vacuum case differs from that in scenarios with matter coupling. To ensure that the resulting effective Hamiltonian constraint remains applicable when matter is coupled, the on-shell condition is not considered not only in this proposition but also throughout the other derivations in this paper.}:
\begin{proposition}\label{eq:SBB}
The totally constrained Hamiltonian theory with the constraint algebra \eqref{eq:effconstraintalgebra} is covariant with respect to the metric $g_{\rho\sigma}^{(\mu)}$ given in \eqref{eq:effg}, namely equation $$\delta g_{\rho\sigma}^{(\mu)}=\mathcal L_{\alpha\mathfrak N+\beta} g_{\rho\sigma}^{(\mu)}$$
holds for all smeared functions $\alpha$ and smeared vector fields $\beta^x\partial_x$ if and only if 
\begin{itemize}
\item[(i)] $\heff$ is independent of $\partial_x^nK_1$ for all $n\geq 1$;
\item[(ii)] The following equation is satisfied for all phase space independent functions $\alpha$ and $N$: 
\begin{equation}\label{eq:mucond}
\alpha\left\{\mu S,\heff[N]\right\}=\left\{\mu S,\heff[\alpha N]\right\}.
\end{equation}
\end{itemize}
\end{proposition}
Notably, the derivation of this proposition relies solely on the constraint algebra \eqref{eq:effconstraintalgebra} and the equations of motion \eqref{eq:effHamton2}. Consequently, the proposition can be easily generalized to models with matter coupling by replacing $\heff$ with the effective total Hamiltonian constraint.

\section{Covariance equations}\label{sec:covarianceequation}
While Proposition~\ref{eq:SBB} provides a necessary and sufficient condition for covariance, it remains unclear whether an effective Hamiltonian constraint $\heff$ satisfying this condition exists and, if so, whether it is unique. In this section, we will show in details that under several physical assumptions, $\heff$ is determined by a mass function $M_{\rm eff}$ satisfying some differential equations, along with an arbitrarily chosen function $\mathcal{R}$. 

Since the Hamiltonian constraint is  a scalar density of weight $1$, it can be expressed in the form of $\heff=E^2F,$
 where $F$ is a scalar field and can be written as a function of basic scalars given by $K_I$, $E^I$ and their derivatives. As discussed in \cite{Zhang:2024khj}, by excluding the derivatives of $K_2$ in the expression of $F$, the condition (i) and the behaviors of the the Poisson  bracket between $\heff[N_1]$ and $\heff[N_2]$ as shown in Eq.~\eqref{eq:effconstraintalgebra} restrict the basic scalars to the following choices:
\begin{equation}
\begin{aligned}
&s_1=E^1, s_2=K_2, s_3=\frac{K_1}{E^2}, s_4=\frac{\partial_xE^1}{E^2},\\
& s_5=\frac{\partial_xs_4}{E^2}, s_6=\frac{\partial_xE^1}{K_1}, s_7=\frac{\partial_xs_4}{K_1}.
\end{aligned}
\end{equation}
In addition, as pointed in \cite{Zhang:2024khj}, in the Schwarzschild solutions of classical GR, there exist $3+1$ decompositions such that $K_I$ or  the derivatives of $E^I$ vanish throughout the entire $t$-slices. Thus the basic scalars $s_6$ and $s_7$ would be ill-defined for such cases. Hence we also exclude $s_6$ and $s_7$ in the expression of $F$ in order to accommodate solutions analogous to the Schwarzschild solutions. 

To summarize, $\heff$ now can be written as
\begin{equation}\label{eq:arbitraryHeff}
\heff=E^2F(\vec s),
\end{equation}
where $\vec s$ denotes the set of $\{s_a\,|\,a=1,2,\cdots,5\}$. Then, we can compute the Poisson bracket
\begin{equation}\label{eq:poissonHH1}
\begin{aligned}
&\{\heff[N_1],\heff[N_2]\}\\
=&\iint \dd x\dd y\big[N_1(x)N_2(y)-N_1(y)N_2(x)\big]\\
&\qquad \times F(\vec s(x))  E^2(y)\left\{E^2(x),F(\vec s(y))\right\}\\
&+\iint \dd x\dd y N_1(x)N_2(y)E^2(x)E^2(y)\\
&\qquad \times \{F(\vec s(x)),F(\vec s(y))\}.
\end{aligned}
\end{equation}
Since $\partial_xK_I$ are not involved in $s_a$, we get
\begin{equation}
\left\{E^2(x),F(\vec s(y))\right\}\propto \delta(x,y),
\end{equation}
which implies that the first term in Eq.~\eqref{eq:poissonHH1} vanishes. Thus, we have
\begin{equation}\label{eq:poissonHHN12general}
\begin{aligned}
&\{\heff[N_1],\heff[N_2]\}\\
=&\sum_{\substack{a,b=1\\a<b}}^5\iint \dd x\dd y\big[N_1(x)N_2(y)-N_1(y)N_2(x)\big]\\
&\times E^2(x)E^2(y)F_a(\vec s(x))F_b(\vec s(y))\{s_a(x),s_b(y)\},
\end{aligned}
\end{equation}
where $\{s_a(x),s_a(y)\}=0$ are applied, and we introduced the convention
\begin{equation}
F_a\equiv \frac{\partial F}{\partial s_a}.
\end{equation} 
 For $a\neq b$, each term in Eq.~\eqref{eq:poissonHHN12general} takes the form
\begin{equation}
\begin{aligned}
B_{ab}=&\iint\dd x\dd y \big[N_1(x)N_2(y)-N_1(y)N_2(x)\big]\\
&\times \mathcal P_a(x)\mathcal P_b(y)\{s_a(x),s_b(y)\}, 
\end{aligned}
\end{equation}
with 
\begin{equation}
\mathcal P_a\equiv E^2F_a.
\end{equation}
Denoting $S_{ab}(x,y)\equiv \{s_a(x),s_b(y)\}$, we have
\begin{equation}
\begin{aligned}
S_{ab}(x,y)\propto\delta(x,y),
\end{aligned}
\end{equation}
for all $$(a,b)\in \{(1,2),(1,3),(1,4),(1,5),(2,3),(2,4),(4,5)\}\equiv V,$$
and hence 
\begin{equation}
B_{ab}=0,\quad \forall\ (a,b)\in V.
\end{equation}
The components of $B_{ab}$ for $(a,b)\notin V$ can be calculated as follows:
\begin{itemize}
\item[(1)] for $(a,b)=(2,5)$,
\begin{equation}
\begin{aligned}
B_{25}=&G\int\dd x(N_1\partial_xN_2-N_2\partial_xN_1)F_2F_5s_4;
\end{aligned}
\end{equation}
\item[(2)] for $(a,b)=(3,4)$,
\begin{equation}
\begin{aligned}
B_{34}=&-2G\int\dd x(N_1\partial_xN_2-N_2\partial_xN_1)F_3F_4;
\end{aligned}
\end{equation}
\item[(3)] for $(a,b)=(3,5)$, 
\begin{equation}
\begin{aligned}
B_{35}=&2G\int\dd x(N_1\partial_xN_2-N_2\partial_xN_1)\frac{1}{E^2}\Big(F_3 \partial_xF_5-F_5\partial_xF_3\Big).
\end{aligned}
\end{equation}
\end{itemize}
Substituting all the above results into Eq.~\eqref{eq:poissonHHN12general}, we get
\begin{equation}\label{eq:HN12general3}
\begin{aligned}
&\{\heff[N_1],\heff[N_2]\}\\
=&2G\int \dd x\left(N_1\partial_xN_2-N_2\partial_xN_1\right)\Bigg\{F_2F_5\frac{s_4}{2}-\\
&\qquad F_3F_4+\sum_{a}\left(F_3\partial_{s_a}F_5-F_5\partial_{s_a}F_3 \right)\frac{\partial_xs_a}{E^2}
  \Bigg\}.
\end{aligned}
\end{equation}
Inserting this equation  into the last equation in Eq.~\eqref{eq:effconstraintalgebra} and comparing the coefficients of $\partial_x K_2$ on its both sides, we get \begin{equation}\label{eq:covariance1p}
\begin{aligned}
\frac{\mu s_1}{2G^2}=F_3\partial_{s_2}F_5-F_5\partial_{s_2}F_3,
\end{aligned}
\end{equation}
and 
\begin{equation}\label{eq:covariant2pp}
\begin{aligned}
\frac{-\mu s_1 s_3s_4}{4G^2}=&\frac{1}{2}F_2F_5s_4-F_3F_4\\
&+ ( \partial_{s_1}F_5 F_3-F_5\partial_{s_1}F_3)s_4\\
&+( \partial_{s_3}F_5 F_3-F_5\partial_{s_3}F_3)\frac{\partial_x s_3}{E^2}\\
&+( \partial_{s_4}F_5 F_3-F_5\partial_{s_4}F_3)s_5\\
&+( \partial_{s_5}F_5 F_3-F_5\partial_{s_5}F_3)\frac{\partial_xs_5}{E^2},
\end{aligned}
\end{equation}
where we used $\partial_xs_1=E^2s_4$ and $\partial_xs_4=E^2s_5$. Now we solve these two equations in aid of the condition (ii).  According to Eq.~\eqref{eq:covariance1p},  $\mu$ depends only on the basic scalars $s_a$, and hence it does not contain $\partial_x s_5/E^2$ or $\partial_x s_3/E^2$. Therefore, Eq.~\eqref{eq:covariant2pp} can be simplified as 
\begin{equation}\label{eq:covariant2ppp}
\begin{aligned}
 0=& F_3\partial_{s_5}F_5-F_5\partial_{s_5}F_3,\\
 0=& F_3 \partial_{s_3}F_5-F_5\partial_{s_3}F_3,\\
0=&\frac{\mu s_1 s_3s_4}{4G^2}+\frac{1}{2}F_2F_5s_4-F_3F_4\\
&+(F_3 \partial_{s_1}F_5 -F_5\partial_{s_1}F_3)s_4\\
&+( F_3\partial_{s_4}F_5 -F_5\partial_{s_4}F_3)s_5.
\end{aligned}
\end{equation}
The first two equations in Eq.~\eqref{eq:covariant2ppp} imply 
\begin{equation}\label{eq:Fsol1}
F=X\left(s_1,s_2,s_4,s_3+C_1(s_1,s_2,s_4)s_5\right),
\end{equation}
for some functions $X(x,y,w,z)$ and $C_1(s_1,s_2,s_4)$. Inserting Eq.~\eqref{eq:Fsol1} into Eq.~\eqref{eq:covariance1p}, we have
\begin{equation}\label{eq:muvalueppp}
\mu=\frac{2 G^2 }{s_1}\partial_{s_2}C_1(\partial_{z}X)^2.
\end{equation}
Hence one can obtain the following identities:
\begin{equation}\label{eq:four20}
\begin{aligned}
&\mu_3F_5-\mu_5F_3=0,\\
&\mu_3\partial_{s_5}F_5-\mu_5\partial_{s_5}F_3=0,\\
&\mu_3\partial_{s_3}F_5-\mu_5\partial_{s_3}F_3=0,
\end{aligned}
\end{equation}
where $\mu_a$ denotes
\begin{equation}
\mu_a=\frac{\partial\mu}{\partial s_a}.
\end{equation}

 As $\heff$ is independent of derivatives of $K_I$, Eq.~\eqref{eq:mucond} can be simplified as
\begin{equation}\label{eq:mucondp}
\alpha\left\{\mu ,\heff[N]\right\}=\left\{\mu ,\heff[\alpha N]\right\}.
\end{equation}
Since $\mu$ also depends on $s_a$ with $a=1,2,\cdots,5$, the Poisson brackets in Eq.~\eqref{eq:mucondp} can be calculated to get
\begin{equation}
\begin{aligned}
&\{\mu,\heff[\alpha N]\}-\alpha\{\mu,\heff[N]\}\\
=&GN\frac{\partial_x\alpha}{E^2}\Bigg[\mu_2F_5s_4-2 \mu_3F_4-6\mu_3F_5\frac{\partial_xE^2}{(E^2)^2}\\
&+\frac{4\mu_3}{N(E^2)^2}\frac{\partial}{\partial x}\left(NE^2F_5\right)-2\mu_4F_3+\mu_5F_2s_4\\
&+6\mu_5F_3\frac{\partial_xE^2}{(E^2)^2}-\frac{4\mu_5}{N(E^2)^2}\frac{\partial}{\partial x}\left(NE^2F_3\right)\Bigg]\\
&+2 GN \frac{\partial_x^2\alpha}{(E^2)^2}\left(\mu_3F_5-\mu_5F_3\right).
\end{aligned}
\end{equation}
Then, Eq.~\eqref{eq:mucondp} leads to
\begin{equation}\label{eq:eqformu}
\begin{aligned}
0=&\mu_3F_5-\mu_5F_3,\\
0=&\mu_2F_5s_4-2 \mu_3F_4-2\mu_4F_3+\mu_5F_2s_4\\
&+4\sum_{a=1}^5\left(\mu_3\partial_{s_a}F_5 -\mu_5\partial_{s_a}F_3\right)\frac{\partial_xs_a}{E^2}.
\end{aligned}
\end{equation}
By Eq.~\eqref{eq:four20}, the last equation in \eqref{eq:eqformu} can be simplified as
\begin{equation}\label{eq:eqformu2}
\begin{aligned}
0=&\mu_2F_5s_4-2 \mu_3F_4-2\mu_4F_3+\mu_5F_2s_4\\
&+4\left(\mu_3\partial_{s_1}F_5-\mu_5\partial_{s_1}F_3\right)s_4\\
&+4\left(\mu_3\partial_{s_4}F_5-\mu_5\partial_{s_4}F_3\right)s_5\\
&+4\left(\mu_3\partial_{s_2}F_5-\mu_5\partial_{s_2}F_3\right)\frac{\partial_xK_2}{E^2}.
\end{aligned}
\end{equation}
Note that only the last term in Eq. \eqref{eq:eqformu2} involves $\partial_xK_2$,  which should vanish.  Using Eqs.~\eqref{eq:Fsol1} and \eqref{eq:muvalueppp} and vanishing the last term in Eq.~\eqref{eq:eqformu2}, we get
\begin{equation}
\begin{aligned}
\frac{4 G^2 (\partial_{s_2}C_1)^2 (\partial_zX)^2 \partial_z^2X}{s_1}=0.
\end{aligned}
\end{equation}
Since neither $\partial_{s_2}C$ nor $\partial_zX$ vanishes in the classical expression of the Hamiltonian constraint $H$, to include $H$ as a special case of the resulting effective Hamiltonian, we have to require 
\begin{equation}
\partial_{z}^2X=0.
\end{equation}
As a result, $F$ takes the form 
\begin{equation}\label{eq:FA}
F=A(s_1,s_2,s_4)+F_3(s_1,s_2,s_4)s_3+F_5(s_1,s_2,s_4)s_5. 
\end{equation}
Taking into account of Eq.~\eqref{eq:covariance1p}, Eq.~\eqref{eq:FA} implies
\begin{equation}
\mu_3=0=\mu_5.
\end{equation}
Then, Eq.~\eqref{eq:eqformu2} becomes 
\begin{equation}\label{eq:eqformu3}
\begin{aligned}
0=&\mu_2F_5s_4-2\mu_4F_3.
\end{aligned}
\end{equation}

We now return to the last equation of Eq.~\eqref{eq:covariant2ppp}. Due to Eq.~\eqref{eq:FA}, it becomes 
\begin{equation}\label{eq:covariant2ppp1}
\begin{aligned}
0=&\frac{1}{2}(\partial_{s_2}A)F_5s_4-F_3(\partial_{s_4}A)\\
&+\left[\frac{\mu s_1 s_4}{4G^2}+\frac{1}{2}(\partial_{s_2}F_3)F_5s_4-F_3\partial_{s_4}F_3\right]s_3\\
&+F_3 (\partial_{s_1}F_5)s_4-F_5 (\partial_{s_1}F_3)s_4\\
&+\left[\frac{1}{2}(\partial_{s_2}F_5) F_5s_4-F_5 \partial_{s_4}F_3\right]s_5.
\end{aligned}
\end{equation}
Observing that in Eq. \eqref{eq:covariant2ppp1} $K_1$ is only contained in $s_3$, we obtain 
\begin{equation}\label{eq:constraintF361o}
\frac{\mu s_1 s_4}{4G^2}+\frac{1}{2}(\partial_{s_2}F_3)F_5s_4-F_3\partial_{s_4}F_3=0.
\end{equation}
Also, since $\partial_x^2E^1$ is contained only in $s_5$, we have
\begin{equation}\label{eq:constraintF362o}
F_5\left[\frac{1}{2}(\partial_{s_2}F_5) s_4- \partial_{s_4}F_3\right]=0,
\end{equation}
which implies the existence of a function $M_{\rm eff}(s_1,s_2,s_4)$ such that
\begin{equation}\label{eq:F3F6M}
\begin{aligned}
F_3=-\frac{\partial M_{\rm eff}}{\partial s_2},\ F_5=\frac{-2}{s_4}\frac{\partial M_{\rm eff}}{\partial s_4}.
\end{aligned}
\end{equation}
Inserting Eq. \eqref{eq:F3F6M} into \eqref{eq:constraintF361o}, we get
\begin{equation}\label{eq:musol2}
\frac{\mu s_1s_4}{4 G^2}= (\partial_{s_2}M_{\rm eff}) \partial_{s_2}\partial_{s_4} M_{\rm eff}-(\partial_{s_4}M_{\rm eff} )\partial_{s_2}^2M_{\rm eff}.
\end{equation}
Due to Eqs.~\eqref{eq:F3F6M} and \eqref{eq:musol2}, Eq.~\eqref{eq:covariant2ppp1} becomes
\begin{equation}\label{eq:QAeq}
\begin{aligned}
0=&-(\partial_{s_4}M_{\rm eff}) \partial_{s_2} (A+2\partial_{s_1}M_{\rm eff})\\
&+(\partial_{s_2}M_{\rm eff})\partial_{s_4}(A+2\partial_{s_1}M_{\rm eff}),
\end{aligned}
\end{equation}
which leads to
\begin{equation}\label{eq:AC2}
\begin{aligned}
A=-2\mathcal R(s_1,M_{\rm eff})-2\partial_{s_1}M_{\rm eff},
\end{aligned}
\end{equation}
for  an arbitrary function $\mathcal R$. Substituting Eq.~\eqref{eq:F3F6M} into Eq.~\eqref{eq:eqformu3}, we get
\begin{equation}\label{eq:solmu3}
(\partial_{s_2}\mu) \partial_{s_4}M_{\rm eff}-(\partial_{s_4}\mu) \partial_{s_2}M_{\rm eff}=0.
\end{equation}
Substituting Eqs.~\eqref{eq:F3F6M} and \eqref{eq:AC2} into Eq.~\eqref{eq:FA}, we finally get expression of the effective Hamiltonian as
\begin{equation}\label{eq:covariance1}
\heff=-2E^2\left[\partial_{s_1}M_{\rm eff}+\frac{\partial_{s_2}M_{\rm eff}}{2}s_3+\frac{\partial_{s_4}M_{\rm eff}}{s_4}s_5+\mathcal R\right],
\end{equation}
where $\mathcal R(s_1,M_{\rm eff})$ could be an arbitrary function, and $M_{\rm eff}(s_1,s_2,s_4)$ {is any solution to} Eqs. \eqref{eq:musol2} and \eqref{eq:solmu3}, which are referred to as the covariance equations. 
%i.e., 
%{\color{red}\begin{equation}\label{eq:covariance2}
%\begin{aligned}
%&\frac{\mu  s_1 s_4}{4G^2}=(\partial_{s_2}M_{\rm eff})\partial_{s_2}\partial_{s_4}M_{\rm eff}-(\partial_{s_4}M_{\rm eff})\partial_{s_2}^2M_{\rm eff},\\
%&(\partial_{s_2}\mu) \partial_{s_4}M_{\rm eff}-(\partial_{s_4}\mu) \partial_{s_2}M_{\rm eff}=0.
%\end{aligned}
%\end{equation}}
Equation \eqref{eq:covariance1} can also be written as
\begin{equation}\label{eq:Qheff}
\heff-2\frac{G\partial_{s_2}M_{\rm eff}}{\partial_xE^1} H_x=-\frac{2}{s_4}\mathcal R\partial_x s_1-\frac{2}{s_4}\partial_xM_{\rm eff},
\end{equation}
by applying $E^2s_3=K_1$ and $E^2s_5=\partial_xs_4$. This implies that in the vacuum case, if $\mathcal R=0$, $M_{\rm eff}$ is a Dirac observable, which actually represents the effective mass of the resulting BH. 

\section{Solutions to the covariance equations}\label{sec:threesolutions}
The effective Hamiltonian constraints of the models can be obtained by solving the covariance equations. As expected,  the classical mass given by
\begin{equation}\label{eq:classicalM}
M_{\rm cl}=\frac{\sqrt{s_1}}{2G}\left[1+(s_2)^2-\frac{(s_4)^2}{4}\right]
\end{equation}
is a solution to Eqs.~\eqref{eq:musol2} and \eqref{eq:solmu3} with $\mu=1$. This verifies the covariance of the classical model. The two solutions proposed in \cite{Zhang:2024khj} are given by
\begin{equation*}
\begin{aligned}
M_{\rm eff}^{(1)}=&\frac{g(s_1)}{2G}+\mathcal F(s_1)\Big[\frac{\sqrt{s_1}}{2G\lambda(s_1)^2}\sin^2\Big(\lambda(s_1)[s_2+\psi(s_1)]\Big)\\
&-\frac{\sqrt{s_1}(s_4)^2}{8G} e^{2i\lambda(s_1)[s_2+\psi(s_1)]}\Big],
\end{aligned}
\end{equation*}
and 
\begin{equation*}
\begin{aligned}
M_{\rm eff}^{(2)}=&\frac{g(s_1)}{2G}+\mathcal F(s_1)\Big[\frac{\sqrt{s_1}}{2 G \lambda(s_1)^2}\sin ^2\left(\lambda(s_1)[s_2+\psi(s_1)]\right)\\
&-\frac{\sqrt{s_1} s_4^2}{8 G} \cos ^2\left(\lambda(s_1)[s_2+\psi(s_1)]\right)\Big],
\end{aligned}
\end{equation*}
where  the arbitrary functions $g$, $\mathcal F$, $\lambda$ and $\psi$ are involved as the integration constants for  the covariance equations. Note that these integration constants were fixed by adopting the $\bar{\mu}$-scheme within the context of loop quantum BH models, and the properties of the resulting spacetimes were studied in \cite{Zhang:2024khj}. However, for a general treatment of these two solutions, the integration constants can be free functions. 
The expressions of the effective Hamiltonian constraints and the resulting metrics in the general treatment are given in Appendix \ref{app:unfixfunction}. It should be noted that the effective Hamiltonian constraint obtained in \cite{Alonso-Bardaji:2023vtl,Belfaqih:2024vfk} can be identified to the one related to $M_{\rm eff}^{(2)}$ by a particular choice of the integration functions. In what follows, we will propose a new solution to the covariance equations as well as its resulting metric.

The new solution is given by 
\begin{equation}\label{eq:meff3ppppp}
\begin{aligned}
M_{\rm eff}^{(3)}=&h(s_1)+\frac{\mathcal F(s_1)\sqrt{s_1}}{2G\lambda(s_1)}\\
&\times \sin(\lambda(s_1)\left[1+(s_2)^2-\frac{(s_4)^2}{4}+\psi(s_1)\right])
\end{aligned}
\end{equation} 
for arbitrary functions $\mathcal F$, $h$, $\lambda$ and $\psi$. Plugging $M_{\rm eff}^{(3)}$ into Eqs.~\eqref{eq:musol2} and \eqref{eq:solmu3}, we get
\begin{equation}\label{eq:mu3}
\mu\equiv \mu_3=\mathcal F(s_1)^2\left\{1-\left[\frac{2G\lambda(s_1)[M_{\rm eff}^{(3)}-h(s_1)]}{\sqrt{s_1}}\right]^2\right\}.
\end{equation}
The resulting effective Hamiltonian constraint reads
\begin{equation}\label{eq:meff3pppppp}
\begin{aligned}
H_{\rm eff}^{(3)}=&-\mathcal F(E^1)\Bigg[\frac{\sqrt{E^1}E^2}{G\lambda(E^1)}\frac{\partial \mathfrak F}{\partial E^1}+\frac{\sqrt{E^1} K_1 K_2}{G}\\
&-\frac{\sqrt{E^1} }{2 G }\partial_x\left(\frac{\partial_xE^1}{E^2}\right)\Bigg]\cos \left(\mathfrak F\right)\\
&-\frac{E^2}{G}\frac{\partial}{\partial E^1}\left(\frac{\sqrt{E^1}\mathcal F(E^1)}{\lambda(E^1)}\right)\sin \left(\mathfrak F\right)\\
&+E^2\left(\frac{\partial h}{\partial E^1}+\mathcal R(E^1,M_{\rm eff}^{(3)})\right),
\end{aligned}
\end{equation}
where 
\begin{equation}
\mathfrak F\equiv \lambda (E^1) \left[1+(K_2)^2-\frac{(\partial_xE^1)^2}{4 (E^2)^2}+\psi (E^1)\right],
\end{equation}
and hence
\begin{equation}
\begin{aligned}
\frac{\partial \mathfrak F}{\partial E^1}= &\lambda'(E^1)\Bigg[1+(K_2)^2-\frac{ (\partial_xE^1)^2 }{4 (E^2)^2}+ \psi (E^1)\Bigg]\\
&+\psi '(E^1)\lambda(E^1).
\end{aligned}
\end{equation}

\subsection{The spacetime metric from $H_{\rm eff}^{(3)}$ with $\mathcal R=0$}
To get the metric, it is convenient to choose the areal gauge 
\begin{equation}\label{eq:arealgauge1}
E^1(x)=x^2.
\end{equation}
Then, solving the diffeomorphism constraint, we get
\begin{equation}\label{eq:arealgauge2}
K_1(x)=\frac{E^2(x)\partial_xK_2(x)}{x}. 
\end{equation}
By the gauge fixing  conditions \eqref{eq:arealgauge1} and \eqref{eq:arealgauge2}, the effective Hamiltonian constraint $H_{\rm eff}^{(3)}$ becomes
\begin{equation}
H_{\rm eff}^{(3)}(x)=-\frac{E^2(x)}{x}\partial_x\hat M_{\rm eff}^{(3)}(x),
\end{equation}
where
\begin{equation*}
\hat M_{\rm eff}^{(3)}(x)=h(x^2)+\frac{x\mathcal F(x^2)\sin(\lambda(x^2)F(x))}{2G\lambda(x^2)},
\end{equation*}
with
\begin{equation}\label{eq:Fmode3}
F(x)=1+K_2(x)^2-\frac{x^2}{E^2(x)^2}+\psi(x^2)
\end{equation}
 is just the value of $M_{\rm eff}^{(3)}$ under the gauge fixing conditions \eqref{eq:arealgauge1} and \eqref{eq:arealgauge2}.

Taking account of the stationary condition
\begin{equation}
\begin{aligned}
0=&\{E^1(x),H_{\rm eff}^{(3)}[N]+H_x[N^x]\},\\
0=&\{E^2(x),H_{\rm eff}^{(3)}[N]+H_x[N^x]\},
\end{aligned}
\end{equation}
we have
\begin{equation}
\begin{aligned}
N(x)&=\frac{x}{E^2(x)},\\
\frac{N^x(x)}{N(x)}&=-\cos \left(\lambda(x^2) F(x)\right) K_2(x)\mathcal F(x^2).
\end{aligned}
\end{equation}
In addition, the vanishing of $H_{\rm eff}^{(3)}$ yields
\begin{equation}
\sin(\lambda(x^2)F(x))=\frac{2G\lambda(x^2)}{x\mathcal F(x^2)}(M-h(x^2)).
\end{equation}
To get the metric in the Schwarzschild-like coordinate, we choose $K_2(x)=0$ so that $N^x(x)=0$. Then, we have
\begin{equation}
\begin{aligned}
&\frac{x^2}{E^2(x)^2}=f_3^{(n)}(x)\\
\equiv &1-\frac{n\pi+(-1)^n\arcsin(\frac{2G\lambda(x^2)(M-h(x^2))}{x\mathcal F(x^2)})}{\lambda(x^2)}+\psi(x^2),
\end{aligned}
\end{equation}
where $n\in \mathbb Z$ is an arbitrary integer. 
We can also get
\begin{equation}
\mu_3=\mathcal F(x^2)^2-\frac{4G^2\lambda(x^2)^2 }{x^2}\left(M-h(x^2)\right)^2.
\end{equation}
Thus, line element in this case reads
\begin{equation}\label{eq:metric3Sch}
\dd s_{(3)}^2=-f_3^{(n)}\dd t^2+\mu_3^{-1}\big(f_3^{(n)}\big)^{-1}\dd x^2+x^2\dd\Omega^2.
\end{equation}

To get the line element of the metric in the Painlev\'e-Gullstrand-like coordinates, one needs to choose the gauge fixing condition $N(x)=1$, i.e., $E^2(x)=x$ (see Appendix \ref{app:unfixfunction} for more details). This condition, 
together with the constraint equation of $H_{\rm eff}^{(3)}$, leads to
\begin{equation}
\begin{aligned}
N^x=\pm \sqrt{\mu_3\big(1-f_3^{(n)}\big)}.
\end{aligned}
\end{equation}
Thus, the line element in the  Painlev\'e-Gullstrand-like coordinates, denoted by $(t_\pm,x,\theta,\phi)$, reads
\begin{equation}\label{eq:metric3PG}
\begin{aligned}
\dd s_{(3)}^2=&-\dd t_\pm^2+\frac{1}{\mu_3}\left(\dd x\pm \sqrt{\mu_3\big(1-f_3^{(n)}\big)}\dd t_\pm \right)^2\\
&+x^2\dd\Omega^2.
\end{aligned}
\end{equation}
where the sign of $+$ corresponds to the ingoing Painlev\'e-Gullstrand-like coordinates and the sign of $-$ corresponds to the outgoing Painlev\'e-Gullstrand-like coordinates. 
It is easy to check that the metric \eqref{eq:metric3Sch} is the same as that of Eq.~\eqref{eq:metric3PG} up to a coordinate transformation, as a consequence of the fact that the theory described by $H_{\rm eff}^{(3)}$ is covariant.

\section{Spacetime structure of $\dd s_{(3)}^2$ in a concrete example}\label{sec:structureds3}
It should be noted that the arguments of the sine function in  $H_{\rm eff}^{(3)}$  involve  $K_2(x)^2$. Inspired by the effective loop quantum black hole models, where the arguments of the sine function typically include  $K_2(x)$, we set  $\lambda(s_1)$  as the square of the corresponding term in the $\bar{\mu}$-scheme of 
those models \cite{Husain:2022gwp,Zhang:2024khj}. Thus, we will consider the spacetime structure of $\dd s_{(3)}^2$ with $\lambda$ and $\psi$ chosen as
\begin{equation}\label{eq:lambdapsi3}
\lambda(s_1)=\frac{\zeta^2}{s_1},\quad \psi(s_1)=0.
\end{equation}
For simplicity,  the other two free functions are chosen as 
\begin{equation}\label{eq:lambdapsi4}
\mathcal F(s_1)=1,\quad h(s_1)=0. 
\end{equation}
Substituting Eqs.~\eqref{eq:lambdapsi3} and \eqref{eq:lambdapsi4}  into Eq.~\eqref{eq:meff3pppppp}, we get the specific expression of $H_{\rm eff}^{(3)}$ as
\begin{equation*}
\begin{aligned}
H_{\rm eff}^{(3)}=&-\Bigg[-\frac{E^2}{G\sqrt{E^1}}\Bigg(1+(K_2)^2-\frac{ (\partial_xE^1)^2 }{4 (E^2)^2}\Bigg)\\
&+\frac{\sqrt{E^1} K_1 K_2}{G}-\frac{\sqrt{E^1} }{2 G }\partial_x\left(\frac{\partial_xE^1}{E^2}\right)\Bigg]\\
&\times\cos \left( \frac{\zeta^2}{E^1} \left[1+(K_2)^2-\frac{(\partial_xE^1)^2}{4 (E^2)^2}\right]\right)\\
&-\frac{3\sqrt{E^1}E^2}{2G\zeta^2}\sin \left( \frac{\zeta^2}{E^1} \left[1+(K_2)^2-\frac{(\partial_xE^1)^2}{4 (E^2)^2}\right]\right)
\end{aligned}
\end{equation*}

The line element of the metric in the Schwarzschild-like coordinates becomes
\begin{equation}\label{eq:schwarzschildds3}
\dd s_{(3)}^2=-\bar f_3^{(n)}\dd t^2+\bar\mu_3^{-1}\big(\bar f_3^{(n)}\big)^{-1}\dd x^2+x^2\dd\Omega^2,
\end{equation} 
with
\begin{equation}\label{eq:f3nexpression}
\begin{aligned}
\bar f_3^{(n)}(x)=&1-(-1)^n\frac{x^2}{\zeta^2}\arcsin(\frac{2GM\zeta^2 }{x^3})-\frac{n\pi x^2}{\zeta^2},\\
\bar\mu_3(x)=&1-\frac{4G^2\zeta^4 M^2}{x^6}.
\end{aligned}
\end{equation}
The metric can return to the Schwarzschild metric as $x$ approaches $\infty$ only  if $n=0$. Hence we set $n=0$ initially to analyze the metric.

To ensure that the argument of the arcsine function remains within its valid domain, one needs to impose the condition\begin{equation}
x\geq \left(2GM\zeta^2\right)^{1/3}\equiv x_{\rm min}.
\end{equation}
For $x$ within this range, we get
\begin{equation}
0\leq \bar\mu_3\leq 1.
\end{equation}
The number of horizons in the spacetime of $\dd s_{(3)}^2$ is determined by the  number of real roots of $\bar f_3^{(0)}$. Observing that as $x\to\infty$, $\bar f_3^{(0)}$ approaches $1$, a positive number, the number of roots of $\bar f_3^{(0)}$ is therefore determined by the sign of its value at $x=\left(2GM\zeta^2\right)^{1/3}$, which reads 
\begin{equation}\label{eq:f302GM}
\mathring{\bar f}_3^{(0)}=1-\frac{\pi}{2}\left(\frac{2GM}{\zeta}\right)^{2/3}.
\end{equation}
Let us introduce 
\begin{equation}
m_o=\frac{\zeta}{2}\left(\frac{2}{\pi}\right)^{3/2}.
\end{equation}
It is easy to see from Eq.~\eqref{eq:f302GM} that there is  no real root of $\bar f_3^{(0)}$ for $GM<m_o$, while there is  one real root of $\bar f_3^{(0)}$ for $GM\geq m_o$.
Now we can explore the spacetime structure of $\dd s_{(3)}^2$ case by case. 
\begin{figure}[t!]
\centering
\begin{subfigure}
\centering
\includegraphics[width=0.5\textwidth]{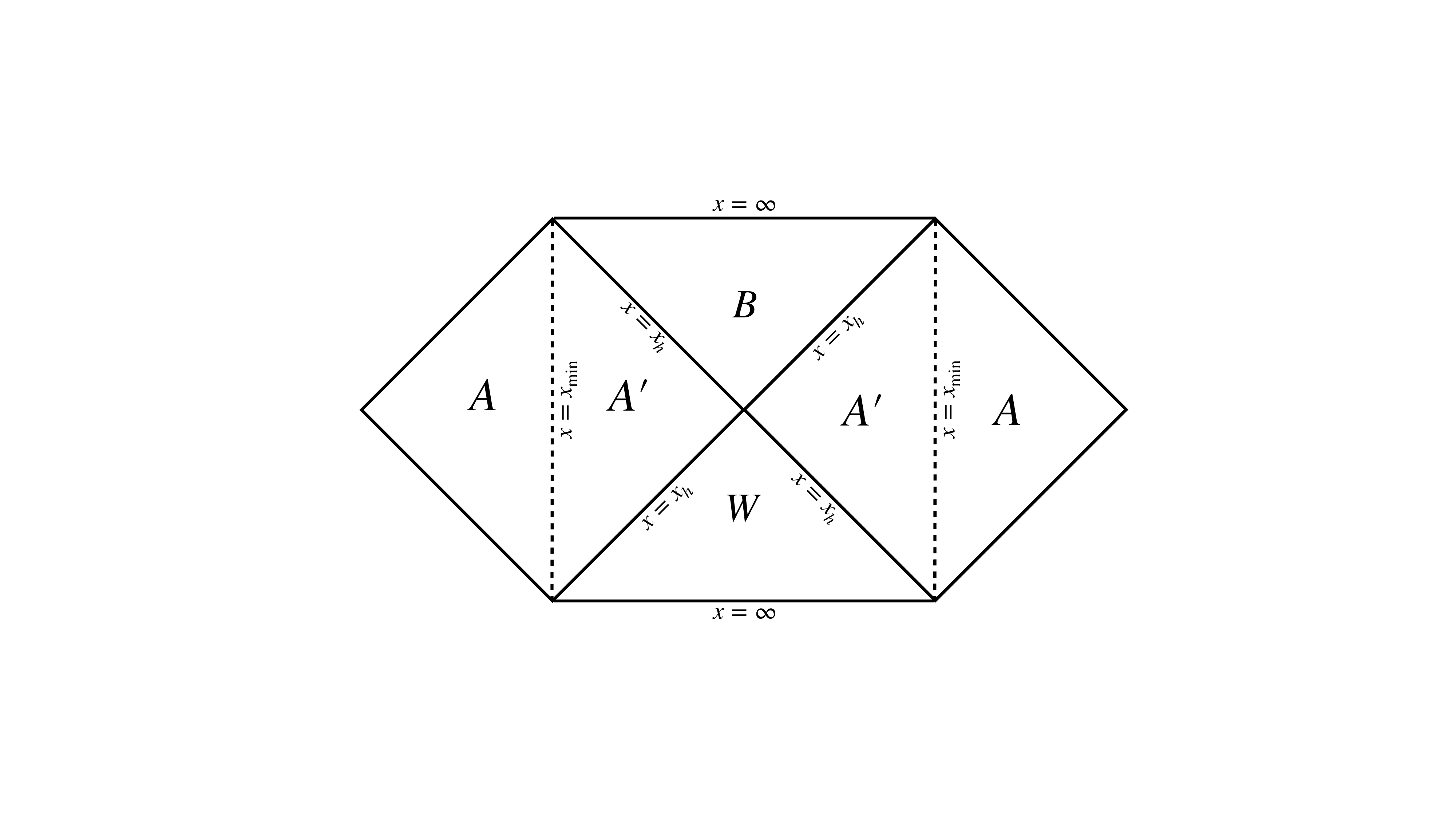}
\caption{The Penrose diagram of the spacetime $\dd s_{(3)}^2$ for $GM<m_o$. The diagram contains the wormhole region $A\cup A'$ with throat occurring at $x=x_{\rm min}$, the  BH region $B$ and the WH region $W$. \newline
}\label{fig:penroseds31}
\end{subfigure}
\begin{subfigure}
\centering
\includegraphics[width=0.5\textwidth]{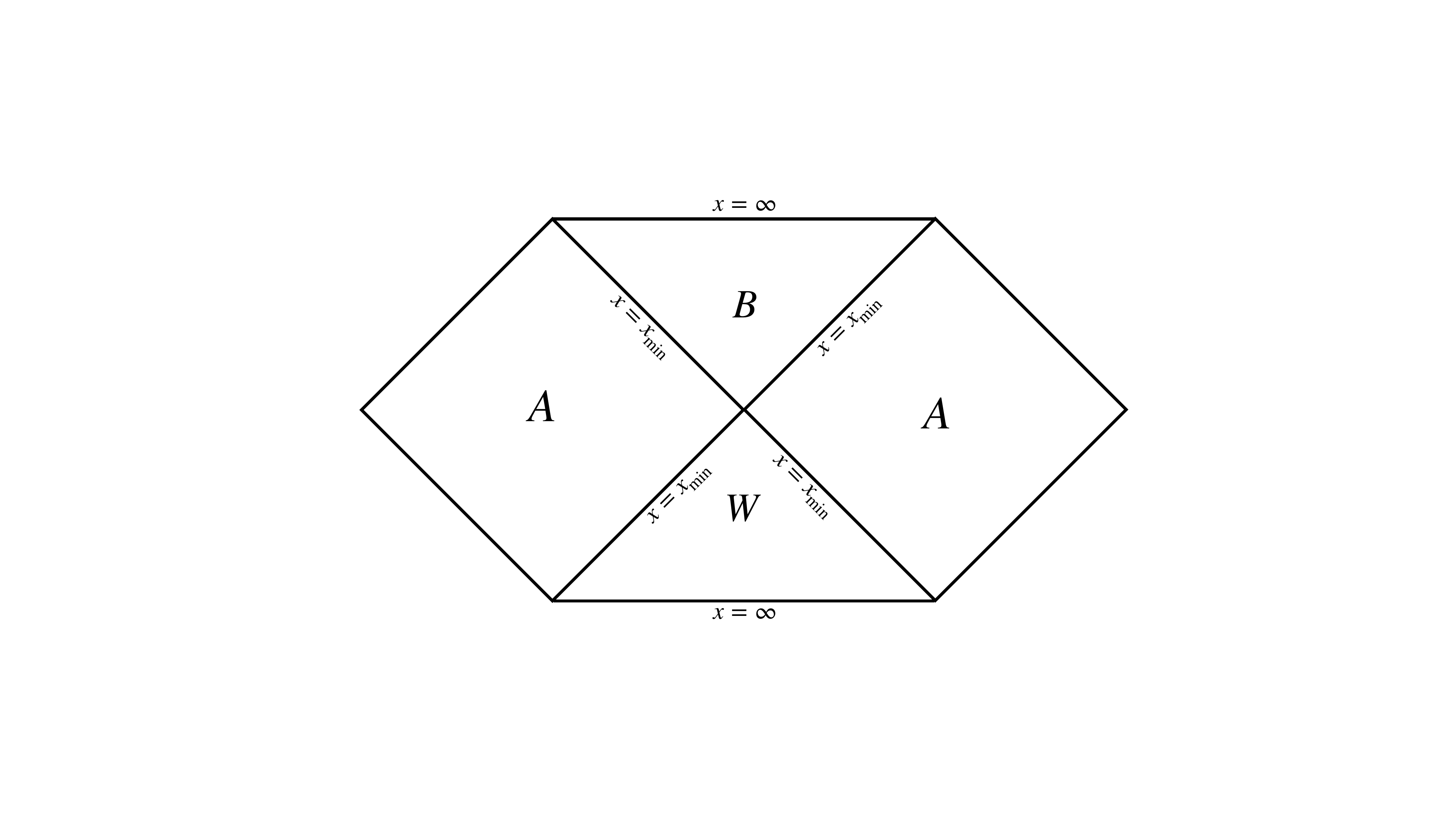}
\caption{The Penrose diagram of the spacetime $\dd s_{(3)}^2$ for $GM=m_o$. The diagram contains the asymptotically flat regions $A$, the BH region $B$ and the WH region $W$.
\newline
}\label{fig:penroseds32}
\end{subfigure}
\begin{subfigure}
\centering
\includegraphics[width=0.5\textwidth]{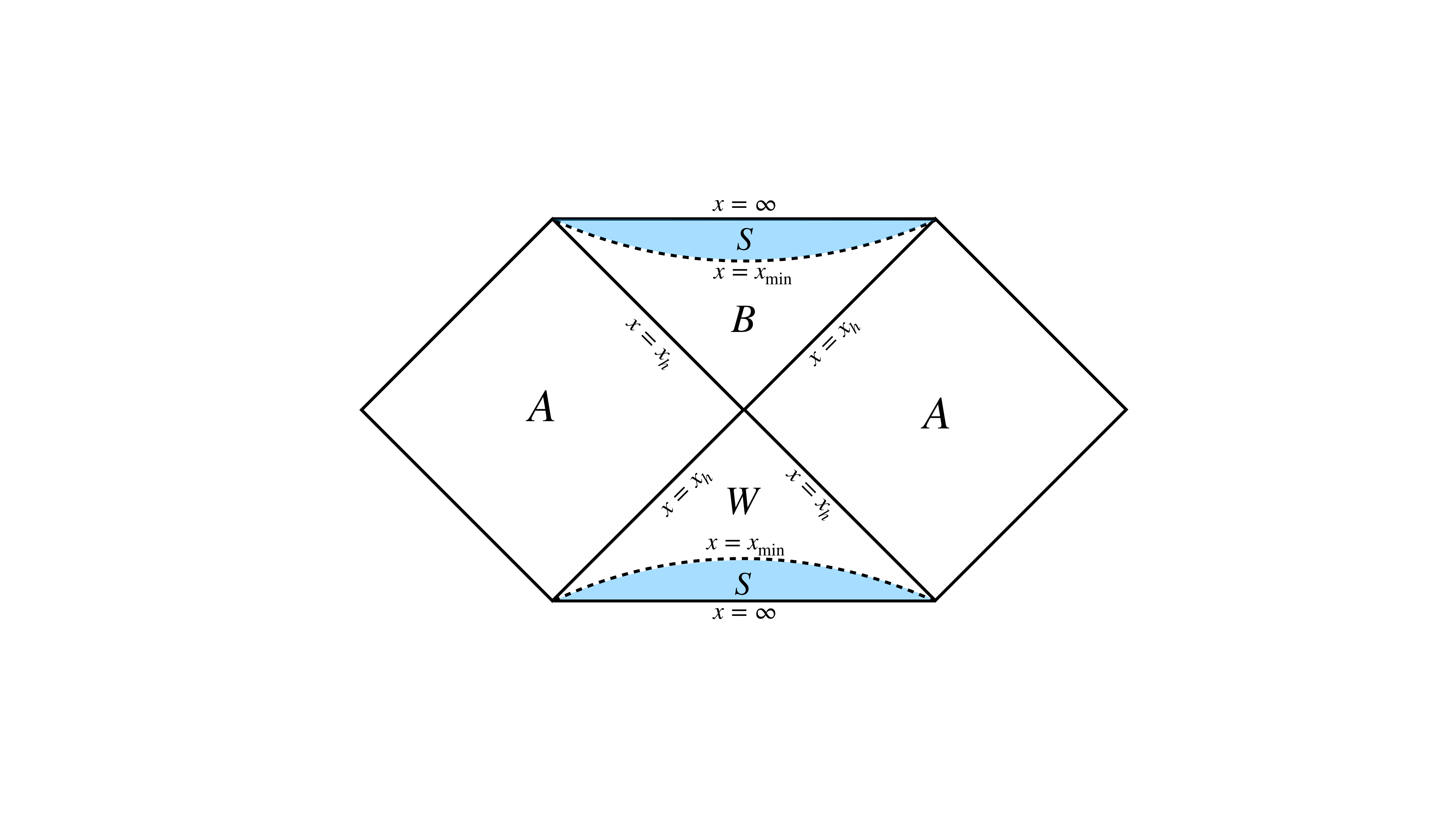}
\caption{The Penrose diagram of the spacetime $\dd s_{(3)}^2$ for $GM>m_o$. In our convention, the shaded regions $S$ are not considered as part of either $B$ or $W$.
The diagram contains asymptotically flat regions $A$, the BH region $B$, the WH region $W$, and the Schwarzschild-de Sitter-like regions $S$.
}\label{fig:penroseds33}
\end{subfigure}
\end{figure}

\subsection{Spacetime structure for $GM<m_o$}

In the case of $GM<m_o$, the coordinates employed in Eq. \eqref{eq:schwarzschildds3} are well-defined for the region of $x>x_{\rm min}$. To extend the spacetime beyond $x=x_{\rm min}$, we introduce a new coordinate $z(x)$, as suggested in~\cite{Alonso-Bardaji:2023niu} for a similar context,  defined by
\begin{equation}\label{eq:determinez}
\left(\frac{\dd z}{\dd x}\right)^2=\mu_3^{-1}. 
\end{equation}
In the coordinates $(t,z,\theta, \phi)$, the metric reads
\begin{equation}\label{eq:newmetricinz}
\dd s_{(3)}^2=-\bar f_3^{(0)}(z)\dd t^2+\big(\bar f_3^{(0)}(z)\big)^{-1}\dd z^2+x(z)^2\dd\Omega^2,
\end{equation}
where $\bar f_3^{(0)}(z):=\bar f_3^{(0)}(x(z))$ and $x(z)$ denotes the inverse function of $z(x)$ determined by Eq.~\eqref{eq:determinez}. The metric \eqref{eq:newmetricinz} is well defined at $x(z)=x_{\rm min}$. Since $x = x_{\rm min}$ is a root of $\mu_3(x)$, the function $x(z)$ exhibits a turning point at $x(z) = x_{\rm min}$. This implies that the extension beyond $x = x_{\rm min}$ turns back to $x>x_{\rm min}$. However, it should be noted that, due to the arcsine function in $\bar f_3^{(n)}$,
after the turning point, the line element should transition to the branch of the arcsine function corresponding to $n = 1$. Then the metric, after the turning point, becomes
\begin{equation}\label{eq:metricIII}
\begin{aligned}
\dd s^2_{(3)}= &-\bar f_3^{(1)}(z)\dd t^2+\big(\bar f_3^{(1)}(z)\big)^{-1}\dd z^2+x(z)^2\dd\Omega^2.
\end{aligned}
 \end{equation}
The choice in Eq.~\eqref{eq:metricIII} ensures the smoothness of the metric, i.e.,
\begin{equation}
 \lim_{z\to z_o^+}\frac{\dd^n}{\dd z^n}f_{\rm eff,III}^{(0)}=\lim_{z\to z_o^-}\frac{\dd^n}{\dd z^n}f_{\rm eff,III}^{(1)}.
\end{equation}

It is obvious that, for $GM<m_o$, $\bar f_3^{(1)}$ has a real root, denoted by $x_h$, satisfying the inequality $x_h>x_{\rm min}$. Hence, there is a horizon in the spacetime region associated with  $\bar f_3^{(1)}$. The expression of $\bar f_3^{(1)}$ in Eq.~\eqref{eq:f3nexpression} indicates that, as $x\to \infty$, the spacetime region associated with  $\bar f_3^{(1)}$ asymptotically approaches to a de-Sitter spacetime. Thus, by gluing the spacetime regions associated with $\bar f_3^{(0)}$ and $\bar f_3^{(1)}$, we get  the Penrose diagram of the spacetime $\dd s_{(3)}^2$ as shown  in Fig. \ref{fig:penroseds31}, where $A$ is the region associated with $\bar f_3^{(0)}$, and $A'\cup B$ and $A'\cup W$ are those associated with $\bar f_3^{(1)}$.

For the region $B\cup A'$ or $W\cup A'$, the Painlev\'e-Gullstrand-like coordinates, denoted by $(t_\pm,x,\theta,\phi)$, can be employed to cover it. The line element in the region $B\cup A'$ is 
\begin{equation}\label{eq:ds3ABW}
\begin{aligned}
\dd s_{(3)}^2=&-\dd t_+^2+\mu_3^{-1}\left(\dd x+ \sqrt{\mu_3\big(1-\bar f_3^{(1)}\big)}\dd t_+ \right)^2\\
&+x^2\dd\Omega^2.
\end{aligned}
\end{equation}
and in the region $W\cup A'$ is 
\begin{equation}\label{eq:ds3ABW2}
\begin{aligned}
\dd s_{(3)}^2=&-\dd t_-^2+\mu_3^{-1}\left(\dd x- \sqrt{\mu_3\big(1-\bar f_3^{(1)}\big)}\dd t_- \right)^2\\
&+x^2\dd\Omega^2.
\end{aligned}
\end{equation}

Besides the coordinates $(t,z,\theta,\phi)$ introduced in Eq.~\eqref{eq:metricIII},  to cover the entire wormhole region $A\cup A'$, we can also introduce the new coordinates $(t,X,\theta,\phi)$ as follows:
\begin{equation}\label{eq:relationXx}
x^3=\frac{2GM\zeta^2}{\sin(X)},
\end{equation}
and hence
\begin{equation}\label{eq:dXdx}
\dd X=\mp \frac{6 \zeta ^2 G M}{x^4 }\frac{\dd x}{\sqrt{\bar\mu_3}}.
\end{equation}
Thus the metric in the new coordinates becomes
\begin{equation}\label{eq:ds3X}
\begin{aligned}
\dd s_{(3)}^2=&-\Phi(X)\dd t^2+\frac{\Xi(X)^2}{9\Phi(X)\sin^2(X)}   \dd X^2\\
&+\Xi(X)^2\dd\Omega^2,
\end{aligned}
\end{equation}
with 
\begin{equation}\label{eq:FX}
\begin{aligned}
\Xi(X)=&\left[\frac{2GM\zeta^2}{\sin(X)}\right]^{1/3},\quad\\
 \Phi(X)=&1-\left[\frac{2GM}{\zeta \sin(X)}\right]^{2/3}X.
\end{aligned}
\end{equation}
The wormhole throat is located at $X=\pi/2$, and the horizon appears at $X=X_o$, where $X_o$, greater than $\pi/2$, is the root of $\Phi(X)$. The range of $X$ reads $0<X<X_o$.

For the cases of $GM\geq m_o$, the procedure to construct the Penrose diagram of the corresponding metrics is similar to the case of $GM<m_o$. Hence, in the following cases, we will introduce the results directly. 

\subsection{Spacetime structure for $GM=m_o$}
In the case of $GM=m_o$, $x=x_{\rm min}=\zeta\sqrt{2/\pi}$ becomes a horizon in the spacetime of $\dd s_{(3)}^2$. The Penrose diagram, as shown in Fig. \ref{fig:penroseds32}, includes the asymptotically flat regions $A$, the BH region $B$ and the WH region $W$. It is convenient to choose the Painlev\'e-Gullstrand-like coordinates $(t_\pm,x,\theta,\phi)$  to cover any of the regions $A$, $B$ and $W$ individually. Then the line element in the regions $A$ reads
\begin{equation}\label{eq:pgmo1}
\begin{aligned}
\dd s_{(3)}^2=&-\dd t^2+\frac{1}{\bar \mu_3}\left(\dd x+\sqrt{\bar \mu_3(1-\bar f_3^{(0)})}\dd t\right)^2\\
&+x^2\dd\Omega^2,
\end{aligned}
\end{equation} 
and the line element in the regions $B$ and $W$ reads
\begin{equation}\label{eq:phmo2}
\begin{aligned}
\dd s_{(3)}^2=&-\dd t_\pm^2+\frac{1}{\bar \mu_3}\left(\dd x+\sqrt{\bar \mu_3(1-\bar f_3^{(0)})}\dd t_\pm\right)^2\\
&+x^2\dd\Omega^2,
\end{aligned}
\end{equation} 
where $t_+$ and $t_-$ are the time coordinate in the region $B$ and $W$ respectively. 

Due to the vanishing of $\mu$ at $x=x_{\rm min}$, the Painlev\'e-Gullstrand-like coordinates $(t_\pm,x,\theta,\phi)$ cannot cross the horizon. To 
avoid the coordinate singularity, we can introduce the coordinate $z$ defined by Eq. \eqref{eq:determinez} or the coordinate $X$ defined by Eq. \eqref{eq:relationXx} to replace the $x$ coordinate in the Painlev\'e-Gullstrand-like coordinates. With the coordinates $(t_\pm,z,\theta,\phi)$ covering $A\cup B$ or $A\cup W$, the line elements respectively read
\begin{equation}\label{eq:zform0}
\begin{aligned}
\dd s_{(3)}^2=&-\dd t_\pm^2+\left(\dd z+\sqrt{\bar \mu_3(z)(1-\bar f_3^{(0)}(z))}\dd t_\pm\right)^2\\
&+x(z)^2\dd\Omega^2.
\end{aligned}\end{equation}
With the  coordinates $(t_\pm,X,\theta,\phi)$, the line elements become
\begin{equation}\label{eq:PGinX}
\begin{aligned}
\dd s_{(3)}^2=&-\dd t_\pm^2+\Xi(X)^2 \left[\frac{\dd X}{3\sin(X)}+\frac{\sqrt{X}}{\zeta}\dd t_\pm\right]^2\\
&+\Xi(X)^2\dd\Omega^2.
\end{aligned}
\end{equation}

\subsection{Spacetime structure for $GM>m_o$}
In this case, $\bar f_3^{(0)}$ has a real root denoted by $x_h$ where  the horizons form. The Penrose diagram, as shown in Fig. \ref{fig:penroseds33}, consists of the asymptotically flat regions $A$, the BH region $B$, the WH region $W$ and the  Schwarzschild-de Sitter-like region $S$. In the current scenario, the surface $x=x_{\rm min}$ occurs inside the horizon and is spacelike. 

In either the region $A\cup B$ or the region $A\cup W$, one can choose the Painlev\'e-Gullstrand-like coordinates $(t_\pm,x,\theta,\phi)$ to cover it. Then, the line element reads 
\begin{equation}
\begin{aligned}
\dd s_{(3)}^2=&-\dd t_\pm^2+\frac{1}{\bar \mu_3}\left(\dd x+\sqrt{\bar \mu_3(1-\bar f_3^{(0)})}\dd t_\pm\right)^2\\
&+x^2\dd\Omega^2,
\end{aligned}
\end{equation}
where as in Eqs. \eqref{eq:ds3ABW} and \eqref{eq:ds3ABW2}, the signs $\pm$ correspond to the regions $A\cup B$ and $A\cup W$ respectively. 

In the shadow region $S$, the Schwarzschild-like coordinate can be chosen, in which the line element reads
\begin{equation}
\dd s_{(3)}^2=-\bar f_3^{(1)}\dd t^2+\bar\mu_3^{-1}(\bar f_3^{(1)})^{-1}\dd x^2+x^2\dd\Omega^2.
\end{equation}  
It is easy to see  that $\bar f_3^{(1)}<0$ for all $x>x_{\rm min}$ in the current case. Moreover, the metric in region $S$ asymptotically approaches the Schwarzschild-de Sitter one with negative mass in the far future.

To define the coordinates covering the entire region $A\cup B\cup S$ or $A\cup W\cup S$, we again use the coordinates $(t_\pm,z,\theta,\phi)$ with $z$ define by Eq. \eqref{eq:determinez}, or the coordinates $(t_\pm,X,\theta,\phi)$ with $X$ given by Eq.~\eqref{eq:relationXx}. Then, the metric in these two coordinate systems will take the same forms as those in Eq. \eqref{eq:zform0} and Eq. \eqref{eq:PGinX} respectively.

\section{Matter coupling}\label{sec:coupledust}
The approach for covariant effective models discussed in previous sections can be extended to include matter coupling, such as a dust fields \cite{PhysRevD.51.5600}. 
In the canonical formulation, the dust fields in the spherically symmetric model can be described by the canonical pairs $(T(x),P_T(x))$ and $(X(x),P_X(x))$, where $P_T$ represents the dust density and $X$ is related to the comoving dust frame. The dust part of the diffeomorphism constraint reads 
\begin{equation}
H_x^{\rm d}=P_T\partial_xT+P_X\partial_x X. 
\end{equation}
Replacing $E^1/(E^2)^2$ in the classical dust Hamiltonian by $\mu E^1/(E^2)^2$, we get a possible expression for the effective dust Hamiltonian as
\begin{equation}\label{eq:effHd}
H_{\rm eff}^{\rm d}=\sqrt{P_T^2+\frac{\mu E^1}{(E^2)^2}(H_x^{\rm d})^2}. 
\end{equation}

Using the fact that $\mu$ is a function of $s_1$ and $M_{\rm eff}$ as indicated by Eq.~\eqref{eq:solmu3}, it is easy to show that 
\begin{equation}
\{H_{\rm eff}^{\rm d}[N_1],H_{\rm eff}^{\rm d}[N_2]\}=H_x^{\rm d}[\mu S(N_1\partial_xN_2-N_2\partial_xN_1)].
\end{equation}
Let $H_x^{\rm tot}=H_x+H_x^{\rm d}$ and $H_{\rm eff}^{\rm tot}=\heff+H_{\rm eff}^{\rm d}$ denote the total diffeomorphism and Hamiltonian constraints, respectively. 
Taking account of Eq.~\eqref{eq:covariance1}, one can further show
\begin{equation}
\{H_{\rm eff}^{\rm d}[N_1],\heff[N_2]\}+\{\heff[N_1],H_{\rm eff}^{\rm d}[N_2]\}=0.
\end{equation}
Consequently, the constraint algebra for $H_{\rm eff}^{\rm tot}$ keeps the same form as given by Eq.~\eqref{eq:effconstraintalgebra}.

Since $H_{\rm eff}^{\rm d}$ does not contain the derivatives of $K_1$, the covariance condition (i) is met by  $H_{\rm eff}^{\rm tot}$. Moreover, a straightforward calculation shows that 
\begin{equation}
\left\{\mu(x),\frac{\mu(y) E^1(y)}{E^2(y)^2}\right\}\propto \delta(x,y).
\end{equation}
As a result, we have
\begin{equation}\label{eq:muhmualpha}
\alpha(x)\{\mu(x), H_{\rm eff}^{\rm d}[N]\}=\{\mu(x),H_{\rm eff}^{\rm d}[\alpha N]\}.
\end{equation}
Furthermore, since the derivatives of $K_2$ is not contained in $H_{\rm eff}^{\rm tot}$ either, Eq.~\eqref{eq:muhmualpha}
implies that $H_{\rm eff}^{\rm tot}$ fulfills the covariance condition (ii).

In the model described by $H_{\rm eff}^{(3)}$, the corresponding value of $\mu\equiv \mu_3$ is given by Eq.~\eqref{eq:mu3}. Thus, the dust Hamiltonian reads,
\begin{equation}
H_{\rm eff}^{{\rm d},3}=\sqrt{P_T^2+\mu_3\frac{E^1}{(E^2)^2}(H_x^{\rm d})^2},
\end{equation}
and thus the total Hamiltonian constraint reads
\begin{equation}
H_{\rm eff}^{{\rm tot},3}=H_{\rm eff}^{(3)}+H_{\rm eff}^{{\rm d},3}. 
\end{equation}

\section{Summary}\label{sec:summary}

The issue of maintaining general covariance in the spherically symmetric effective models of QG has been studied in previous sections. We retain the kinematical variables $(E^I, K_I)$, with $I=1,2$, and the classical form of the diffeomorphism constraint $H_x$, while leave the effective Hamiltonian constraint $\heff$  undetermined to accommodate QG effects. The diffeomorphism constraint  and the effective Hamiltonian constraint are assumed to obey the Poisson relation \eqref{eq:effconstraintalgebra}, where a free factor $\mu$ is introduced to account for QG effects. 

To ensure general covariance, the classical metric is modified into the effective one $g_{\rho\sigma}^{(\mu)}$ by incorporating $\mu$ into its expression, which ensures that the algebra \eqref{eq:effconstraintalgebra} continues to describe hypersurface deformations as in the classical case. Then, by requiring that the gauge transformation of  $g_{\rho\sigma}^{(\mu)}$  coincides with its corresponding diffeomorphism transformation, the sufficient and necessary conditions for the covariance are derived. Based on the covariance conditions, we ultimately arrive at  Eq.~\eqref{eq:covariance1}, which links $\heff$ and $M_{\rm eff}$, with $M_{\rm eff}$ satisfying Eqs.~\eqref{eq:musol2} and \eqref{eq:solmu3}. Note that $\heff$ includes a free function $\mathcal{R}$, and by setting $\mathcal{R} = 0$ $M_{\rm eff}$ becomes a Dirac observable representing the BH mass.

Solving Eqs.~\eqref{eq:musol2} and \eqref{eq:solmu3}, we have obtained three families of $M_{\rm eff}$, denoted by $M_{\rm eff}^{(i)}$ with $i=1,2,3$, which involve some free functions as the constants of integration. Note that in \cite{Zhang:2024khj}, the properties of the models resulting from $M_{\rm eff}^{(1)}$ and $M_{\rm eff}^{(2)}$ were studied with these integration constants fixed by adopting the $\bar{\mu}$-scheme within the context of loop quantum BH models. 
In the current paper, we focus on the new solution $M_{\rm eff}^{(3)}$. In the resulting spacetime, the classical singularity is resolved, and the spacetime is extended beyond the singularity into a Schwarzschild-de Sitter spacetime. Notably, the new quantum-corrected spacetime does not contain any Cauchy horizons, which might imply its stability under perturbations (see, e.g., \cite{Cao:2023aco,Cao:2023par,Lin:2024beb} for further discussion on Cauchy horizon in effective quantum BH models).

It should be noted that,  from the perspective of a Taylor expansion of the sine function in \eqref{eq:meff3ppppp} and \eqref{eq:meff3pppppp}, the appearance of higher-order derivatives of $E^1$ is expected. However, the sine function involving the spatial derivative of $E^1$ does not introduce additional degrees of freedom, as it contains only spatial derivatives and hence does not lead to any ghost fields.  Furthermore, the absence of higher-order derivatives of $E^2$ in our construction is not an arbitrary assumption, but rather a consequence of the requirement of general covariance, which strongly constrains the form of the effective Hamiltonian. Why general covariance permits higher-order derivatives of $E^1$ while excluding them for $E^2$ is an interesting issue that deserves further investigation.

By proposing the effective dust Hamiltonian \eqref{eq:effHd}, the effective model of gravity coupled to the dust fields has been obtained. Thus our covariant approach establishes a fundamental platform for future research involving matter couplings. This would allow for the study of BH formation through the collapse of a dust ball, which is left for our future study. Furthermore, it is desirable to generalize the framework developed in this paper to models with different symmetries, such as the axially symmetric case.

\begin{acknowledgments}
C.Z. acknowledges the valuable discussions with Luca Cafaro, Kristina Giesel, Hongguang Liu, Chun-Yen Lin, Dongxue Qu, Farshid Soltani, Edward Wilson-Ewing, and Stefan Andreas Weigl.  This  work is supported by the NSFC Grant Nos. 12275022 and 12165005, ``the Fundamental Research Funds for the Central Universities'', %(No. 2253100010), 
and the
National Science Centre, Poland as part of the OPUS 24 Grant No. 2022/47/B/ST2/02735.

\end{acknowledgments}

\onecolumngrid

\appendix 

\section{The metrics associated with $M_{\rm eff}^{(1)}$ and  $M_{\rm eff}^{(2)}$}\label{app:unfixfunction}

\subsection{The metric associated with $M_{\rm eff}^{(1)}$}
The first solution, as proposed in \cite{Zhang:2024khj}, is obtained by considering the polymerization of $M_{\rm cl}$ in Eq.~\eqref{eq:classicalM}. We introduce
\begin{equation}
\begin{aligned}
M_{\rm eff}^{(1)}=&\frac{g(s_1)}{2G}+\mathcal F(s_1)\left[\frac{\sqrt{s_1}}{2G\lambda(s_1)^2}\sin^2\Big(\lambda(s_1)[s_2+\psi(s_1)]\Big)-\frac{\sqrt{s_1}(s_4)^2}{8G} \exp\Big(2i\lambda(s_1)[s_2+\psi(s_1)] \Big)\right],
\end{aligned}
\end{equation}
with arbitrary functions $g$, $\mathcal F$, $\lambda$ and $\psi$.  Substituting this expression into Eq.~\eqref{eq:musol2} and \eqref{eq:solmu3}, we have
\begin{equation}
\mu\equiv \mu_1=\mathcal F(s_1)^2.
\end{equation}
The resulting Hamiltonian constraint  reads
\begin{equation}
\begin{aligned}
H_{\rm eff}^{(1)}=&-\frac{\mathcal F(E^1)\sqrt{E^1}}{2G\lambda(E^1)^2}\left[K_1\lambda(E^1)+2 E^2 \frac{\partial \mathfrak P(E^1)}{\partial E^1}\right]\sin (2 \mathfrak P(E^1))\\
&-\frac{E^2}{G}\frac{\partial}{\partial E^1}\left[\frac{\sqrt{E^1}\mathcal F(E^1)}{\lambda(E^1)^2}\right]\sin^2(\mathfrak P(E^1))\\
&+\left[\frac{(\partial_xE^1)^2 }{4 G  E^2}\frac{\partial}{\partial E^1}\Big(\sqrt{E^1}\mathcal F(E^1)\Big)+\frac{\sqrt{E^1}\mathcal F(E^1)}{2G}\partial_x\left(\frac{\partial_xE^1}{ E^2}\right)\right]e^{2 i \mathfrak P(E^1)}\\
&+\frac{i\sqrt{E^1} \mathcal F(E^1)(\partial_xE^1)^2}{2GE^2}\left[\frac{ \lambda (E^1) K_1}{2E^2}+\frac{\partial \mathfrak P(E^1)}{\partial E^1}  \right]e^{2 i \mathfrak P(E^1)}\\
&-\frac{E^2}{G }\frac{\partial g(E^1)}{\partial E^1}+E^2\mathcal R(E^1,M_{\rm eff}^{(1)}),
\end{aligned}
\end{equation}
%\begin{widetext}
%\begin{equation}
%\begin{aligned}
%H_{\rm eff}^{(1)}=&-\frac{\sqrt{E^1} E^2 }{G \lambda (E^1)^2}\left[K_2 \lambda '(E^1) + \psi (E^1) \lambda '(E^1)+ \psi '(E^1)\lambda(E^1)\right]\sin (2 \lambda (E^1) (K_2+\psi (E^1)))\\
%&-\frac{\sqrt{E^1}K_1 \sin (2 \lambda (E^1) (K_2+\psi (E^1)))}{2 G \lambda (E^1)}-\frac{2E^2}{G\lambda (E^1)^2}\left(\frac{1}{4  \sqrt{E^1}}-\frac{\sqrt{E^1}  \lambda '(E^1) }{ \lambda (E^1)}\right)\sin^2( \lambda (E^1) (K_2+\psi (E^1)))\\
%&-\frac{E^2}{2 G \sqrt{E^1}}+\frac{(\partial_xE^1)^2 }{8 G \sqrt{E^1} E^2}e^{2 i \lambda (E^1) (K_2+\psi (E^1))}+\frac{\sqrt{E^1}}{2G}\partial_x\left(\frac{\partial_x^2E^1}{ E^2}\right)e^{2 i \lambda (E^1) (K_2+\psi (E^1))}\\
%&+\frac{i\sqrt{E^1} (\partial_xE^1)^2}{2GE^2}\left(\frac{ \lambda (E^1) K_1}{2E^2}+ \lambda '(E^1)K_2  + \psi (E^1) \lambda '(E^1)+ \lambda (E^1) \psi '(E^1) \right)e^{2 i \lambda (E^1) (K_2+\psi (E^1))}
%\end{aligned}
%\end{equation}
where we introduced the abbreviation 
\begin{equation}\label{eq:polymerK2}
\mathfrak P\equiv \lambda (E^1) \Big[K_2+\psi (E^1)\Big]
\end{equation}
to denote the polymerization of $K_2$. 
%\end{widetext}

%\subsubsection{The spacetime metric from $H_{\rm eff}^{(1)}$ with $\mathcal R=0$}\label{sec:derivativemetric}
%To get the metric, it is convenient to choose the areal gauge 
%\begin{equation}\label{eq:arealgauge1}
%E^1(x)=x^2.
%\end{equation}
%Then, solving the diffeomorphism constraint, we get
%\begin{equation}\label{eq:arealgauge2}
%K_1(x)=\frac{E^2(x)\partial_xK_2(x)}{x}. 
%\end{equation}
In what follows, we focus on the case of $\mathcal R=0$. Substituting the gauge fixing condition \eqref{eq:arealgauge1} and \eqref{eq:arealgauge2} into $H_{\rm eff}^{(1)}$, we get
\begin{equation}
\begin{aligned}
H_{\rm eff}^{(1)}(x)=-\frac{E^2(x)}{x}\partial_x\hat M_{\rm eff}^{(1)}(x),
\end{aligned}
\end{equation}
where $\hat M_{\rm eff}^{(1)}(x)$ denotes $M_{\rm eff}^{(1)}(x)$  with Eqs.~\eqref{eq:arealgauge1} and \eqref{eq:arealgauge2} substituted. 

We are interested in the stationary solution, in which $N$ and $N^x$ should be chosen so that $\dot E^I=0$, i.e., 
\begin{equation}\label{eq:stationarycondition}
\begin{aligned}
0=&\{E^1(x),H_{\rm eff}^{(1)}[N]+H_x[N^x]\},\\
0=&\{E^2(x),H_{\rm eff}^{(1)}[N]+H_x[N^x]\},
\end{aligned}
\end{equation}
leading to
\begin{equation}
\begin{aligned}
N(x)=&\frac{x}{E^2(x)},\\
\frac{N^x(x)}{N(x)}=& \frac{\mathcal F\left(x^2\right)}{2 \lambda \left(x^2\right)E^2(x)^2} \Bigg[-E^2(x)^2\sin \left(2 \lambda \left(x^2\right) \left[K_2(x)+\psi \left(x^2\right)\right]\right)\\
&+2 i x^2 \lambda \left(x^2\right)^2 e^{2 i \lambda \left(x^2\right) \left(K_2(x)+\psi \left(x^2\right)\right)}\Bigg].
\end{aligned}
\end{equation}

To write down the metric in the Schwarzschild-like coordinates $(t,x,\theta,\phi)$, we choose the Schwarzschild gauge such that 
\begin{equation}\label{eq:schwaraschildgauge}
N^x=0.
\end{equation}
This condition, together with the constraint $H_{\rm eff}^{(1)}=0$, results in
\begin{equation}\label{eq:metric1SCH}
\begin{aligned}
E^2(x)&=\pm\frac{x^2 \mathcal F(x^2)}{\sqrt{\left[g\left(x^2\right)-2 G M\right] \left(x f\left(x^2\right)+\lambda \left(x^2\right)^2 \left(g\left(x^2\right)-2 G M\right)\right)}}.
\end{aligned}
\end{equation}
We thus get the metric
\begin{equation}\label{eq:ds1insch}
\begin{aligned}
\dd s_{(1)}^2=-f_1\dd t^2+\mu_1^{-1}f_1^{-1}\dd x^2+x^2\dd\Omega^2,
\end{aligned}
\end{equation}
with
\begin{equation}
f_1(x)=\frac{ g\left(x^2\right)-2 G M}{x \mathcal F\left(x^2\right)}\left[1+\frac{\lambda \left(x^2\right)^2 \left(g\left(x^2\right)-2 G M\right)}{x \mathcal F\left(x^2\right)}\right],
\end{equation}
and 
\begin{equation}
\mu_1(x)=\mathcal F(x^2)^2.
\end{equation}

To write down the  metric in the Painlev\'e-Gullstrand-like coordinates $(t_p,x,\theta,\phi)$, we choose  the  gauge 
\begin{equation}\label{eq:PGgauge}
E^2(x)=x,
\end{equation}
leading to $N=1$. This condition,  together with the  constraint $H_{\rm eff}^{(1)}=0$, gives
\begin{equation}
N^x(x)=\pm  \sqrt{\mu_1(1-f_1(x))}.
\end{equation}
Thus, the metric in the Painlev\'e-Gullstrand-like coordinates $(t_\pm,x,\theta,\phi)$  is
\begin{equation}\label{eq:metric1PG}
\begin{aligned}
\dd s_{(1)}^2=-\dd t_\pm^2+\mu_1^{-1}\left(\dd x\pm \sqrt{\mu_1(1-f_1})\dd t_\pm\right)^2+x^2\dd\Omega^2,
\end{aligned}
\end{equation}
where the sign of $+$ corresponds to the ingoing Painlev\'e-Gullstrand-like coordinates and the sign of $-$ corresponds to the outgoing Painlev\'e-Gullstrand-like coordinates. 
It is easy to check that  the metric \eqref{eq:metric1PG} is the same as the one in Eq.~\eqref{eq:metric1SCH} up to a coordinate transformation. This fact is compatible with fact that the theory described by $H_{\rm eff}^{(1)}$ is covariant.

\subsection{The spacetime metric associated with $M_{\rm eff}^{(2)}$}
The second solution is obtained by choosing 
\begin{equation}
\begin{aligned}
M_{\rm eff}^{(2)}=&\frac{g(s_1)}{2G}+\mathcal F(s_1)\left[\frac{\sqrt{s_1}}{2 G \lambda(s_1)^2}\sin ^2\left(\lambda(s_1)[s_2+\psi(s_1)]\right)-\frac{\sqrt{s_1} s_4^2}{8 G} \cos ^2\left(\lambda(s_1)[s_2+\psi(s_1)]\right)\right],
\end{aligned}
\end{equation}
for arbitrary functions $g$, $\mathcal F$, $\lambda$ and $\psi$. Substituting $M_{\rm eff}^{(2)}$ into Eqs.~\eqref{eq:musol2} and \eqref{eq:solmu3} leads to
\begin{equation}\label{eq:mu2origional}
\mu\equiv \mu_2=\mathcal F(s_1)\left[\mathcal F(s_1)+\frac{\lambda (s_1)^2}{\sqrt{s_1}} \left(g(s_1)-2 G M_{\rm eff}^{(2)}\right)\right].
\end{equation}
The resulting Hamiltonian constraint $H_{\rm eff}^{(2)}$ reads
\begin{equation}
\begin{aligned}
H_{\rm eff}^{(2)}=&-\frac{\mathcal F(E^1)\sqrt{E^1}}{2G\lambda(E^1)^2}\left[K_1\lambda(E^1)+2 E^2 \frac{\partial \mathfrak P(E^1)}{\partial E^1}\right]\sin (2 \mathfrak P(E^1))\\
&-\frac{E^2}{G}\frac{\partial}{\partial E^1}\left[\frac{\sqrt{E^1}\mathcal F(E^1)}{\lambda(E^1)^2}\right]\sin^2(\mathfrak P(E^1))\\
&+\left[\frac{(\partial_xE^1)^2 }{4 G  E^2}\frac{\partial}{\partial E^1}\Big(\sqrt{E^1}\mathcal F(E^1)\Big)+\frac{\sqrt{E^1}\mathcal F(E^1)}{2G}\partial_x\left(\frac{\partial_xE^1}{ E^2}\right)\right]\cos^2( \mathfrak P(E^1))\\
&-\frac{\sqrt{E^1} \mathcal F(E^1)(\partial_xE^1)^2}{4GE^2}\left[\frac{ \lambda (E^1) K_1}{2E^2}+\frac{\partial \mathfrak P(E^1)}{\partial E^1}  \right]\sin(2\mathfrak P(E^1))\\
&-\frac{E^2}{G }\frac{\partial g(E^1)}{\partial E^1}+E^2\mathcal R(E^1,M_{\rm eff}^{(1)}),
\end{aligned}
\end{equation}
with $\mathfrak P$ given by Eq.~\eqref{eq:polymerK2}.  
%\begin{widetext}
%\begin{equation}
%\begin{aligned}
%H_{\rm eff}^{(2)}=&-\frac{\sqrt{E^1}K_1}{2 G \lambda (E^1)} \sin (2 \lambda (E^1) (K_2+\psi (E^1)))+\frac{E^2}{G\lambda(E^1)^2}\left(\frac{2\sqrt{E^1} \lambda '(E^1) }{ \lambda (E^1)}-\frac{1}{2  \sqrt{E^1} }\right)\sin^2( \lambda (E^1) (K_2+\psi (E^1)))\\
%&-\frac{\sqrt{E^1}E^2}{G\lambda(E^1)^2}\left(K_2 \lambda '(E^1) +\psi (E^1) \lambda '(E^1)+ \psi '(E^1) \lambda (E^1)\right)\sin (2 \lambda (E^1) (K_2+\psi (E^1)))\\
%&-\frac{E^2}{2 G \sqrt{E^1}}+\frac{(\partial_xE^1)^2 }{8 G \sqrt{E^1} E^2}\cos^2( \lambda (E^1) (K_2+\psi (E^1)))+\frac{\sqrt{E^1}}{2 G }\partial_x\left(\frac{\partial_xE^1}{E^2}\right)\cos^2( \lambda (E^1) (K_2+\psi (E^1)))\\
%&-\frac{\sqrt{E^1}(\partial_xE^1)^2}{4GE^2}\left(\frac{\lambda (E^1) K_1  }{2E^2}+K_2\lambda '(E^1)+\psi (E^1) \lambda '(E^1) + \lambda (E^1)\psi '(E^1)\right)\sin (2 \lambda (E^1) (K_2+\psi (E^1)))
%\end{aligned}
%\end{equation}
%\end{widetext}

In what follows, we again focus on the case of $\mathcal R=0$.
We still choose the areal gauge to solve the dynamics. Substituting the gauge fixing conditions \eqref{eq:arealgauge1} and \eqref{eq:arealgauge2} into $H_{\rm eff}^{(2)}$, we have
\begin{equation}
H_{\rm eff}^{(2)}(x)=-\frac{E^2(x)}{x}\partial_x\hat M_{\rm eff}^{(2)},
\end{equation}
where $\hat M_{\rm eff}^{(2)}(x)$ is the value of $M_{\rm eff}^{(2)}$ with the conditions \eqref{eq:arealgauge1} and \eqref{eq:arealgauge2} substituted. 

Since the stationary solution is still focused on, we apply the similar procedure as in Eq.~\eqref{eq:stationarycondition}, obtaining 
%\begin{equation}
%\begin{aligned}
%0=&\{E^1(x),H_{\rm eff}^{(2)}[N]+H_x[N^x]\},\\
%0=&\{E^2(x),H_{\rm eff}^{(2)}[N]+H_x[N^x]\},
%\end{aligned}
%\end{equation}
%leading to 
\begin{equation}
\begin{aligned}
N(x)&=\frac{x}{E^2(x)},\\
\frac{N^x(x)}{N(x)}&=-\frac{\mathcal F\left(x^2\right)}{2 E^2(x)^2 \lambda \left(x^2\right)}  \left(E^2(x)^2+x^2 \lambda \left(x^2\right)^2\right) \sin \Big(2 \lambda(x^2) \left[K_2(x)+\psi \left(x^2\right)\right]\Big).
\end{aligned}
\end{equation}

Then, the Schwarzschild gauge \eqref{eq:schwaraschildgauge}, together with the constraint equation $H_{\rm eff}^{(2)}=0$, results in 
%To write the metric in the Schwarzschild-like coordinate, we need to gauge making $N^x(x)=0$. This condition, together with the constraint equation $H_{\rm eff}^{(2)}=0$, results in
\begin{equation}
\begin{aligned}
&E^2(x)=\pm\frac{\sqrt{x}^3\sqrt{\mathcal F(x^2)}}{\sqrt{g(x^2)-2GM}},\\
&\sin \left(\lambda \left(x^2\right) \left[K_2(x)+\psi \left(x^2\right)\right]\right)=0.
\end{aligned}
\end{equation}
Consequently, the  value of $\mu_2$ is
\begin{equation}
\mu_2=\mathcal F(x^2)^2\left(1+\frac{\lambda(x^2)^2\left(g(x^2)-2GM\right)}{x\mathcal F(x^2)}\right).
\end{equation}
Substituting the above results into the expression $g_{\rho\sigma}^{(\mu)}$, we get the metric
\begin{equation}\label{eq:metric2SCH}
\dd  s_{(2)}^2=-f_2\dd t^2+\mu_2^{-1}f_2^{-1}\dd x^2+x^2\dd\Omega^2,
\end{equation}
with
\begin{equation}\label{eq:f2}
f_2=\frac{g\left(x^2\right)-2 G M}{x \mathcal F\left(x^2\right)}.
\end{equation}

To write the metric in the Painlev\'e-Gullstrand-like coordinates, we again need the gauge \eqref{eq:PGgauge} which, together with
%We can also choose the Painlev\'e-Gullstrand-like coordinate to write the metric. This require us to choose the gauge with $N=1$, i.e., $E^2(x)=x$. This condition, together with 
the constraint $H_{\rm eff}^{(2)}=0$, results in
\begin{equation}
N^x(x)=\pm \sqrt{\mu_2(1-f_2)}.
\end{equation}
Thus, the metric in the Painlev\'e-Gullstrand-like coordinates $(t_p,x,\theta,\phi)$ is 
\begin{equation}\label{eq:metric2GP}
\dd s^2=-\dd t_\pm^2+\mu_2^{-1}\left[\dd x\pm \sqrt{\mu_2(1-f_2)}\dd t_\pm\right]^2+x^2\dd\Omega^2,
\end{equation}
where the  sign of $+$ corresponds to the ingoing Painlev\'e-Gullstrand-like coordinates and the sign of $-$ corresponds to the outgoing Painlev\'e-Gullstrand-like coordinates. 
It can be easily checked that the metric \eqref{eq:metric2SCH} is  the same as the  one given in Eq.~\eqref{eq:metric2GP}, as a consequence of the fact that the theory described by $H_{\rm eff}^{(2)}$ is covariant.

%\bibliography{reference}

\begin{thebibliography}{44}
\expandafter\ifx\csname natexlab\endcsname\relax\def\natexlab#1{#1}\fi
\expandafter\ifx\csname bibnamefont\endcsname\relax
  \def\bibnamefont#1{#1}\fi
\expandafter\ifx\csname bibfnamefont\endcsname\relax
  \def\bibfnamefont#1{#1}\fi
\expandafter\ifx\csname citenamefont\endcsname\relax
  \def\citenamefont#1{#1}\fi
\expandafter\ifx\csname url\endcsname\relax
  \def\url#1{\texttt{#1}}\fi
\expandafter\ifx\csname urlprefix\endcsname\relax\def\urlprefix{URL }\fi
\providecommand{\bibinfo}[2]{#2}
\providecommand{\eprint}[2][]{\url{#2}}

\bibitem[{\citenamefont{Ashtekar et~al.}(2015)\citenamefont{Ashtekar, Berger,
  Isenberg, and MacCallum}}]{ashtekar2015general}
\bibinfo{author}{\bibfnamefont{A.}~\bibnamefont{Ashtekar}},
  \bibinfo{author}{\bibfnamefont{B.~K.} \bibnamefont{Berger}},
  \bibinfo{author}{\bibfnamefont{J.}~\bibnamefont{Isenberg}}, \bibnamefont{and}
  \bibinfo{author}{\bibfnamefont{M.}~\bibnamefont{MacCallum}},
  \emph{\bibinfo{title}{General relativity and gravitation: a centennial
  perspective}} (\bibinfo{publisher}{Cambridge University Press},
  \bibinfo{year}{2015}).

\bibitem[{\citenamefont{Penrose}(1965)}]{PhysRevLett.14.57}
\bibinfo{author}{\bibfnamefont{R.}~\bibnamefont{Penrose}},
  \bibinfo{journal}{Phys. Rev. Lett.} \textbf{\bibinfo{volume}{14}},
  \bibinfo{pages}{57} (\bibinfo{year}{1965}),
  \urlprefix\url{https://link.aps.org/doi/10.1103/PhysRevLett.14.57}.

\bibitem[{\citenamefont{Rovelli and Smolin}(1988)}]{Rovelli:1987df}
\bibinfo{author}{\bibfnamefont{C.}~\bibnamefont{Rovelli}} \bibnamefont{and}
  \bibinfo{author}{\bibfnamefont{L.}~\bibnamefont{Smolin}},
  \bibinfo{journal}{Phys. Rev. Lett.} \textbf{\bibinfo{volume}{61}},
  \bibinfo{pages}{1155} (\bibinfo{year}{1988}).

\bibitem[{\citenamefont{Polchinski}(1998)}]{polchinski1998string}
\bibinfo{author}{\bibfnamefont{J.~G.} \bibnamefont{Polchinski}},
  \emph{\bibinfo{title}{String theory, volume I: An introduction to the bosonic
  string}} (\bibinfo{publisher}{Cambridge university press Cambridge},
  \bibinfo{year}{1998}).

\bibitem[{\citenamefont{Ambjorn et~al.}(2001)\citenamefont{Ambjorn, Jurkiewicz,
  and Loll}}]{Ambjorn:2001cv}
\bibinfo{author}{\bibfnamefont{J.}~\bibnamefont{Ambjorn}},
  \bibinfo{author}{\bibfnamefont{J.}~\bibnamefont{Jurkiewicz}},
  \bibnamefont{and} \bibinfo{author}{\bibfnamefont{R.}~\bibnamefont{Loll}},
  \bibinfo{journal}{Nucl. Phys. B} \textbf{\bibinfo{volume}{610}},
  \bibinfo{pages}{347} (\bibinfo{year}{2001}), \eprint{hep-th/0105267}.

\bibitem[{\citenamefont{Surya}(2019)}]{Surya:2019ndm}
\bibinfo{author}{\bibfnamefont{S.}~\bibnamefont{Surya}},
  \bibinfo{journal}{Living Rev. Rel.} \textbf{\bibinfo{volume}{22}},
  \bibinfo{pages}{5} (\bibinfo{year}{2019}), \eprint{1903.11544}.

\bibitem[{\citenamefont{Donoghue}(2012)}]{Donoghue:2012zc}
\bibinfo{author}{\bibfnamefont{J.~F.} \bibnamefont{Donoghue}},
  \bibinfo{journal}{AIP Conf. Proc.} \textbf{\bibinfo{volume}{1483}},
  \bibinfo{pages}{73} (\bibinfo{year}{2012}), \eprint{1209.3511}.

\bibitem[{\citenamefont{Jha}(2023)}]{Jha:2022svf}
\bibinfo{author}{\bibfnamefont{R.}~\bibnamefont{Jha}},
  \bibinfo{journal}{SciPost Phys. Lect. Notes} \textbf{\bibinfo{volume}{73}},
  \bibinfo{pages}{1} (\bibinfo{year}{2023}), \eprint{2204.03537}.

\bibitem[{\citenamefont{Tibrewala}(2014)}]{Tibrewala:2013kba}
\bibinfo{author}{\bibfnamefont{R.}~\bibnamefont{Tibrewala}},
  \bibinfo{journal}{Class. Quant. Grav.} \textbf{\bibinfo{volume}{31}},
  \bibinfo{pages}{055010} (\bibinfo{year}{2014}), \eprint{1311.1297}.

\bibitem[{\citenamefont{Bojowald et~al.}(2015)\citenamefont{Bojowald, Brahma,
  and Reyes}}]{Bojowald:2015zha}
\bibinfo{author}{\bibfnamefont{M.}~\bibnamefont{Bojowald}},
  \bibinfo{author}{\bibfnamefont{S.}~\bibnamefont{Brahma}}, \bibnamefont{and}
  \bibinfo{author}{\bibfnamefont{J.~D.} \bibnamefont{Reyes}},
  \bibinfo{journal}{Phys. Rev. D} \textbf{\bibinfo{volume}{92}},
  \bibinfo{pages}{045043} (\bibinfo{year}{2015}), \eprint{1507.00329}.

\bibitem[{\citenamefont{Wu et~al.}(2018)\citenamefont{Wu, Bojowald, and
  Ma}}]{Wu:2018mhg}
\bibinfo{author}{\bibfnamefont{J.-P.} \bibnamefont{Wu}},
  \bibinfo{author}{\bibfnamefont{M.}~\bibnamefont{Bojowald}}, \bibnamefont{and}
  \bibinfo{author}{\bibfnamefont{Y.}~\bibnamefont{Ma}}, \bibinfo{journal}{Phys.
  Rev. D} \textbf{\bibinfo{volume}{98}}, \bibinfo{pages}{106009}
  (\bibinfo{year}{2018}), \eprint{1809.04465}.

\bibitem[{\citenamefont{Bojowald}(2019)}]{Bojowald:2019dry}
\bibinfo{author}{\bibfnamefont{M.}~\bibnamefont{Bojowald}}
  (\bibinfo{year}{2019}), \eprint{1906.04650}.

\bibitem[{\citenamefont{Bojowald}(2020)}]{Bojowald:2020unm}
\bibinfo{author}{\bibfnamefont{M.}~\bibnamefont{Bojowald}},
  \bibinfo{journal}{Phys. Rev. D} \textbf{\bibinfo{volume}{102}},
  \bibinfo{pages}{046006} (\bibinfo{year}{2020}), \eprint{2007.16066}.

\bibitem[{\citenamefont{Han and Liu}(2024)}]{Han:2022rsx}
\bibinfo{author}{\bibfnamefont{M.}~\bibnamefont{Han}} \bibnamefont{and}
  \bibinfo{author}{\bibfnamefont{H.}~\bibnamefont{Liu}},
  \bibinfo{journal}{Phys. Rev. D} \textbf{\bibinfo{volume}{109}},
  \bibinfo{pages}{084033} (\bibinfo{year}{2024}), \eprint{2212.04605}.

\bibitem[{\citenamefont{Gambini et~al.}(2022)\citenamefont{Gambini, Olmedo, and
  Pullin}}]{Gambini:2022dec}
\bibinfo{author}{\bibfnamefont{R.}~\bibnamefont{Gambini}},
  \bibinfo{author}{\bibfnamefont{J.}~\bibnamefont{Olmedo}}, \bibnamefont{and}
  \bibinfo{author}{\bibfnamefont{J.}~\bibnamefont{Pullin}},
  \bibinfo{journal}{Phys. Rev. D} \textbf{\bibinfo{volume}{105}},
  \bibinfo{pages}{026017} (\bibinfo{year}{2022}), \eprint{2201.01616}.

\bibitem[{\citenamefont{Bojowald}(2022)}]{Bojowald:2022zog}
\bibinfo{author}{\bibfnamefont{M.}~\bibnamefont{Bojowald}},
  \bibinfo{journal}{Phys. Rev. D} \textbf{\bibinfo{volume}{105}},
  \bibinfo{pages}{108901} (\bibinfo{year}{2022}), \eprint{2203.06049}.

\bibitem[{\citenamefont{Ashtekar et~al.}(2023)\citenamefont{Ashtekar, Olmedo,
  and Singh}}]{Ashtekar:2023cod}
\bibinfo{author}{\bibfnamefont{A.}~\bibnamefont{Ashtekar}},
  \bibinfo{author}{\bibfnamefont{J.}~\bibnamefont{Olmedo}}, \bibnamefont{and}
  \bibinfo{author}{\bibfnamefont{P.}~\bibnamefont{Singh}}
  (\bibinfo{year}{2023}), \eprint{2301.01309}.

\bibitem[{\citenamefont{Giesel et~al.}(2024{\natexlab{a}})\citenamefont{Giesel,
  Liu, Singh, and Weigl}}]{Giesel:2023hys}
\bibinfo{author}{\bibfnamefont{K.}~\bibnamefont{Giesel}},
  \bibinfo{author}{\bibfnamefont{H.}~\bibnamefont{Liu}},
  \bibinfo{author}{\bibfnamefont{P.}~\bibnamefont{Singh}}, \bibnamefont{and}
  \bibinfo{author}{\bibfnamefont{S.~A.} \bibnamefont{Weigl}},
  \bibinfo{journal}{Phys. Rev. D} \textbf{\bibinfo{volume}{110}},
  \bibinfo{pages}{104016} (\bibinfo{year}{2024}{\natexlab{a}}),
  \eprint{2308.10953}.

\bibitem[{\citenamefont{Bojowald and Duque}(2024)}]{Bojowald:2024beb}
\bibinfo{author}{\bibfnamefont{M.}~\bibnamefont{Bojowald}} \bibnamefont{and}
  \bibinfo{author}{\bibfnamefont{E.~I.} \bibnamefont{Duque}},
  \bibinfo{journal}{Phys. Rev. D} \textbf{\bibinfo{volume}{109}},
  \bibinfo{pages}{084044} (\bibinfo{year}{2024}), \eprint{2401.15040}.

\bibitem[{\citenamefont{Date}(2010)}]{Date:2010xr}
\bibinfo{author}{\bibfnamefont{G.}~\bibnamefont{Date}}, in
  \emph{\bibinfo{booktitle}{{Refresher Course for College Teachers}}}
  (\bibinfo{year}{2010}), \eprint{1010.2062}.

\bibitem[{\citenamefont{Pons et~al.}(1997)\citenamefont{Pons, Salisbury, and
  Shepley}}]{Pons:1996av}
\bibinfo{author}{\bibfnamefont{J.~M.} \bibnamefont{Pons}},
  \bibinfo{author}{\bibfnamefont{D.~C.} \bibnamefont{Salisbury}},
  \bibnamefont{and} \bibinfo{author}{\bibfnamefont{L.~C.}
  \bibnamefont{Shepley}}, \bibinfo{journal}{Phys. Rev. D}
  \textbf{\bibinfo{volume}{55}}, \bibinfo{pages}{658} (\bibinfo{year}{1997}),
  \eprint{gr-qc/9612037}.

\bibitem[{\citenamefont{Bojowald et~al.}(2018)\citenamefont{Bojowald, Brahma,
  and Yeom}}]{Bojowald:2018xxu}
\bibinfo{author}{\bibfnamefont{M.}~\bibnamefont{Bojowald}},
  \bibinfo{author}{\bibfnamefont{S.}~\bibnamefont{Brahma}}, \bibnamefont{and}
  \bibinfo{author}{\bibfnamefont{D.-h.} \bibnamefont{Yeom}},
  \bibinfo{journal}{Phys. Rev. D} \textbf{\bibinfo{volume}{98}},
  \bibinfo{pages}{046015} (\bibinfo{year}{2018}), \eprint{1803.01119}.

\bibitem[{\citenamefont{Lan et~al.}(2023)\citenamefont{Lan, Yang, Guo, and
  Miao}}]{Lan:2023cvz}
\bibinfo{author}{\bibfnamefont{C.}~\bibnamefont{Lan}},
  \bibinfo{author}{\bibfnamefont{H.}~\bibnamefont{Yang}},
  \bibinfo{author}{\bibfnamefont{Y.}~\bibnamefont{Guo}}, \bibnamefont{and}
  \bibinfo{author}{\bibfnamefont{Y.-G.} \bibnamefont{Miao}},
  \bibinfo{journal}{Int. J. Theor. Phys.} \textbf{\bibinfo{volume}{62}},
  \bibinfo{pages}{202} (\bibinfo{year}{2023}), \eprint{2303.11696}.

\bibitem[{\citenamefont{Zhang et~al.}(2022)\citenamefont{Zhang, Ma, Song, and
  Zhang}}]{Zhang:2021wex}
\bibinfo{author}{\bibfnamefont{C.}~\bibnamefont{Zhang}},
  \bibinfo{author}{\bibfnamefont{Y.}~\bibnamefont{Ma}},
  \bibinfo{author}{\bibfnamefont{S.}~\bibnamefont{Song}}, \bibnamefont{and}
  \bibinfo{author}{\bibfnamefont{X.}~\bibnamefont{Zhang}},
  \bibinfo{journal}{Phys. Rev. D} \textbf{\bibinfo{volume}{105}},
  \bibinfo{pages}{024069} (\bibinfo{year}{2022}), \eprint{2107.10579}.

\bibitem[{\citenamefont{Zhang}(2021)}]{Zhang:2021xoa}
\bibinfo{author}{\bibfnamefont{C.}~\bibnamefont{Zhang}},
  \bibinfo{journal}{Phys. Rev. D} \textbf{\bibinfo{volume}{104}},
  \bibinfo{pages}{126003} (\bibinfo{year}{2021}), \eprint{2106.08202}.

\bibitem[{\citenamefont{Husain et~al.}(2022{\natexlab{a}})\citenamefont{Husain,
  Kelly, Santacruz, and Wilson-Ewing}}]{Husain:2021ojz}
\bibinfo{author}{\bibfnamefont{V.}~\bibnamefont{Husain}},
  \bibinfo{author}{\bibfnamefont{J.~G.} \bibnamefont{Kelly}},
  \bibinfo{author}{\bibfnamefont{R.}~\bibnamefont{Santacruz}},
  \bibnamefont{and}
  \bibinfo{author}{\bibfnamefont{E.}~\bibnamefont{Wilson-Ewing}},
  \bibinfo{journal}{Phys. Rev. Lett.} \textbf{\bibinfo{volume}{128}},
  \bibinfo{pages}{121301} (\bibinfo{year}{2022}{\natexlab{a}}),
  \eprint{2109.08667}.

\bibitem[{\citenamefont{Husain et~al.}(2022{\natexlab{b}})\citenamefont{Husain,
  Kelly, Santacruz, and Wilson-Ewing}}]{Husain:2022gwp}
\bibinfo{author}{\bibfnamefont{V.}~\bibnamefont{Husain}},
  \bibinfo{author}{\bibfnamefont{J.~G.} \bibnamefont{Kelly}},
  \bibinfo{author}{\bibfnamefont{R.}~\bibnamefont{Santacruz}},
  \bibnamefont{and}
  \bibinfo{author}{\bibfnamefont{E.}~\bibnamefont{Wilson-Ewing}},
  \bibinfo{journal}{Phys. Rev. D} \textbf{\bibinfo{volume}{106}},
  \bibinfo{pages}{024014} (\bibinfo{year}{2022}{\natexlab{b}}),
  \eprint{2203.04238}.

\bibitem[{\citenamefont{Giesel et~al.}(2024{\natexlab{b}})\citenamefont{Giesel,
  Liu, Rullit, Singh, and Weigl}}]{Giesel:2023tsj}
\bibinfo{author}{\bibfnamefont{K.}~\bibnamefont{Giesel}},
  \bibinfo{author}{\bibfnamefont{H.}~\bibnamefont{Liu}},
  \bibinfo{author}{\bibfnamefont{E.}~\bibnamefont{Rullit}},
  \bibinfo{author}{\bibfnamefont{P.}~\bibnamefont{Singh}}, \bibnamefont{and}
  \bibinfo{author}{\bibfnamefont{S.~A.} \bibnamefont{Weigl}},
  \bibinfo{journal}{Phys. Rev. D} \textbf{\bibinfo{volume}{110}},
  \bibinfo{pages}{104017} (\bibinfo{year}{2024}{\natexlab{b}}),
  \eprint{2308.10949}.

\bibitem[{\citenamefont{Zhang et~al.}(2025)\citenamefont{Zhang, Lewandowski,
  Ma, and Yang}}]{Zhang:2024khj}
\bibinfo{author}{\bibfnamefont{C.}~\bibnamefont{Zhang}},
  \bibinfo{author}{\bibfnamefont{J.}~\bibnamefont{Lewandowski}},
  \bibinfo{author}{\bibfnamefont{Y.}~\bibnamefont{Ma}}, \bibnamefont{and}
  \bibinfo{author}{\bibfnamefont{J.}~\bibnamefont{Yang}},
  \bibinfo{journal}{Phys. Rev. D} \textbf{\bibinfo{volume}{111}},
  \bibinfo{pages}{L081504} (\bibinfo{year}{2025}), \eprint{2407.10168}.

\bibitem[{\citenamefont{Zhang}(2024)}]{numericalResult}
\bibinfo{author}{\bibfnamefont{C.}~\bibnamefont{Zhang}},
  \bibinfo{howpublished}{\url{https://github.com/czhangUW/covarianceBHModels}}
  (\bibinfo{year}{2024}).

\bibitem[{\citenamefont{Bojowald}(2004)}]{Bojowald:2004af}
\bibinfo{author}{\bibfnamefont{M.}~\bibnamefont{Bojowald}},
  \bibinfo{journal}{Class. Quant. Grav.} \textbf{\bibinfo{volume}{21}},
  \bibinfo{pages}{3733} (\bibinfo{year}{2004}), \eprint{gr-qc/0407017}.

\bibitem[{\citenamefont{Bojowald and Swiderski}(2006)}]{Bojowald:2005cb}
\bibinfo{author}{\bibfnamefont{M.}~\bibnamefont{Bojowald}} \bibnamefont{and}
  \bibinfo{author}{\bibfnamefont{R.}~\bibnamefont{Swiderski}},
  \bibinfo{journal}{Class. Quant. Grav.} \textbf{\bibinfo{volume}{23}},
  \bibinfo{pages}{2129} (\bibinfo{year}{2006}), \eprint{gr-qc/0511108}.

\bibitem[{\citenamefont{Gambini et~al.}(2023)\citenamefont{Gambini, Olmedo, and
  Pullin}}]{Gambini:2022hxr}
\bibinfo{author}{\bibfnamefont{R.}~\bibnamefont{Gambini}},
  \bibinfo{author}{\bibfnamefont{J.}~\bibnamefont{Olmedo}}, \bibnamefont{and}
  \bibinfo{author}{\bibfnamefont{J.}~\bibnamefont{Pullin}},
  \emph{\bibinfo{title}{{Quantum Geometry and Black Holes}}}
  (\bibinfo{year}{2023}), \eprint{2211.05621}.

\bibitem[{\citenamefont{Ashtekar}(1986)}]{PhysRevLett.57.2244}
\bibinfo{author}{\bibfnamefont{A.}~\bibnamefont{Ashtekar}},
  \bibinfo{journal}{Phys. Rev. Lett.} \textbf{\bibinfo{volume}{57}},
  \bibinfo{pages}{2244} (\bibinfo{year}{1986}),
  \urlprefix\url{https://link.aps.org/doi/10.1103/PhysRevLett.57.2244}.

\bibitem[{\citenamefont{Barbero~G.}(1995)}]{BarberoG:1994eia}
\bibinfo{author}{\bibfnamefont{J.~F.} \bibnamefont{Barbero~G.}},
  \bibinfo{journal}{Phys. Rev. D} \textbf{\bibinfo{volume}{51}},
  \bibinfo{pages}{5507} (\bibinfo{year}{1995}), \eprint{gr-qc/9410014}.

\bibitem[{\citenamefont{Bojowald and Paily}(2012)}]{Bojowald:2011aa}
\bibinfo{author}{\bibfnamefont{M.}~\bibnamefont{Bojowald}} \bibnamefont{and}
  \bibinfo{author}{\bibfnamefont{G.~M.} \bibnamefont{Paily}},
  \bibinfo{journal}{Phys. Rev. D} \textbf{\bibinfo{volume}{86}},
  \bibinfo{pages}{104018} (\bibinfo{year}{2012}), \eprint{1112.1899}.

\bibitem[{\citenamefont{Bojowald and Duque}(2023)}]{Bojowald:2023xat}
\bibinfo{author}{\bibfnamefont{M.}~\bibnamefont{Bojowald}} \bibnamefont{and}
  \bibinfo{author}{\bibfnamefont{E.~I.} \bibnamefont{Duque}},
  \bibinfo{journal}{Phys. Rev. D} \textbf{\bibinfo{volume}{108}},
  \bibinfo{pages}{084066} (\bibinfo{year}{2023}), \eprint{2310.06798}.

\bibitem[{\citenamefont{Alonso-Bardaji and
  Brizuela}(2024)}]{Alonso-Bardaji:2023vtl}
\bibinfo{author}{\bibfnamefont{A.}~\bibnamefont{Alonso-Bardaji}}
  \bibnamefont{and} \bibinfo{author}{\bibfnamefont{D.}~\bibnamefont{Brizuela}},
  \bibinfo{journal}{Phys. Rev. D} \textbf{\bibinfo{volume}{109}},
  \bibinfo{pages}{044065} (\bibinfo{year}{2024}), \eprint{2310.12951}.

\bibitem[{\citenamefont{Belfaqih et~al.}(2024)\citenamefont{Belfaqih, Bojowald,
  Brahma, and Duque}}]{Belfaqih:2024vfk}
\bibinfo{author}{\bibfnamefont{I.~H.} \bibnamefont{Belfaqih}},
  \bibinfo{author}{\bibfnamefont{M.}~\bibnamefont{Bojowald}},
  \bibinfo{author}{\bibfnamefont{S.}~\bibnamefont{Brahma}}, \bibnamefont{and}
  \bibinfo{author}{\bibfnamefont{E.~I.} \bibnamefont{Duque}}
  (\bibinfo{year}{2024}), \eprint{2407.12087}.

\bibitem[{\citenamefont{Alonso-Bardaji
  et~al.}(2023)\citenamefont{Alonso-Bardaji, Brizuela, and
  Vera}}]{Alonso-Bardaji:2023niu}
\bibinfo{author}{\bibfnamefont{A.}~\bibnamefont{Alonso-Bardaji}},
  \bibinfo{author}{\bibfnamefont{D.}~\bibnamefont{Brizuela}}, \bibnamefont{and}
  \bibinfo{author}{\bibfnamefont{R.}~\bibnamefont{Vera}},
  \bibinfo{journal}{Phys. Rev. D} \textbf{\bibinfo{volume}{107}},
  \bibinfo{pages}{064067} (\bibinfo{year}{2023}), \eprint{2302.10619}.

\bibitem[{\citenamefont{Brown and Kucha\ifmmode~\check{r}\else
  \v{r}\fi{}}(1995)}]{PhysRevD.51.5600}
\bibinfo{author}{\bibfnamefont{J.~D.} \bibnamefont{Brown}} \bibnamefont{and}
  \bibinfo{author}{\bibfnamefont{K.~V.}
  \bibnamefont{Kucha\ifmmode~\check{r}\else \v{r}\fi{}}},
  \bibinfo{journal}{Phys. Rev. D} \textbf{\bibinfo{volume}{51}},
  \bibinfo{pages}{5600} (\bibinfo{year}{1995}),
  \urlprefix\url{https://link.aps.org/doi/10.1103/PhysRevD.51.5600}.

\bibitem[{\citenamefont{Cao et~al.}(2024{\natexlab{a}})\citenamefont{Cao, Li,
  Wu, and Zhou}}]{Cao:2023aco}
\bibinfo{author}{\bibfnamefont{L.-M.} \bibnamefont{Cao}},
  \bibinfo{author}{\bibfnamefont{L.-Y.} \bibnamefont{Li}},
  \bibinfo{author}{\bibfnamefont{L.-B.} \bibnamefont{Wu}}, \bibnamefont{and}
  \bibinfo{author}{\bibfnamefont{Y.-S.} \bibnamefont{Zhou}},
  \bibinfo{journal}{Eur. Phys. J. C} \textbf{\bibinfo{volume}{84}},
  \bibinfo{pages}{507} (\bibinfo{year}{2024}{\natexlab{a}}),
  \eprint{2308.10746}.

\bibitem[{\citenamefont{Cao et~al.}(2024{\natexlab{b}})\citenamefont{Cao, Li,
  Liu, and Zhou}}]{Cao:2023par}
\bibinfo{author}{\bibfnamefont{L.-M.} \bibnamefont{Cao}},
  \bibinfo{author}{\bibfnamefont{L.-Y.} \bibnamefont{Li}},
  \bibinfo{author}{\bibfnamefont{X.-Y.} \bibnamefont{Liu}}, \bibnamefont{and}
  \bibinfo{author}{\bibfnamefont{Y.-S.} \bibnamefont{Zhou}},
  \bibinfo{journal}{Phys. Rev. D} \textbf{\bibinfo{volume}{109}},
  \bibinfo{pages}{064083} (\bibinfo{year}{2024}{\natexlab{b}}),
  \eprint{2312.04301}.

\bibitem[{\citenamefont{Lin et~al.}(2025)\citenamefont{Lin, Zhang, and
  Bravo-Gaete}}]{Lin:2024beb}
\bibinfo{author}{\bibfnamefont{J.}~\bibnamefont{Lin}},
  \bibinfo{author}{\bibfnamefont{X.}~\bibnamefont{Zhang}}, \bibnamefont{and}
  \bibinfo{author}{\bibfnamefont{M.}~\bibnamefont{Bravo-Gaete}},
  \bibinfo{journal}{Phys. Rev. D} \textbf{\bibinfo{volume}{111}},
  \bibinfo{pages}{106025} (\bibinfo{year}{2025}), \eprint{2412.01448}.

\end{thebibliography}

\end{document}